\documentclass[12pt]{article}
\usepackage{amsthm}
\usepackage{latexsym}
\usepackage{amssymb}
\usepackage{amsmath}
\usepackage{mathrsfs}
\usepackage{euscript}

\newcommand{\R}{\mathbb{R}}
\newcommand{\N}{\mathbb{N}}

\newcommand{\p}{\partial}
\renewcommand{\d}{\mathrm{d}}

\newcommand{\scri}{{\mathscr I}}

\newcommand{\hook}{{\setlength{\unitlength}{11pt}   
                   \begin{picture}(.833,.8)
                   \put(.15,.08){\line(1,0){.35}}
                   \put(.5,.08){\line(0,1){.5}}
                   \end{picture}}}
\newtheorem{definition}{Definition}[section]
\newtheorem{theorem}{Theorem}
\newtheorem{proposition}{Proposition}[section]
\newtheorem{corollary}{Corollary}[section]
\newtheorem{lemma}{Lemma}[section]
\newtheorem{remark}{Remark}[section]
\begin{document}\mbox{}

\vspace{0.25in}

\begin{center}
\huge{Regularity at space-like and null infinity}

\vspace{0.25in}

\large{Lionel J. MASON\footnote{The Mathematical Institute, 24-29 St
    Giles', OXFORD OX1 3LB, UNITED KINGDOM. \\ \noindent
    lmason@maths.ox.ac.uk} \&
    Jean-Philippe NICOLAS\footnote{M.A.B., Institut de Math\'ematiques
    de Bordeaux, Universit\'e Bordeaux 1, 351 cours de la
    Lib\'eration, 33405 TALENCE cedex,
    FRANCE. \\ \noindent Jean-Philippe.Nicolas@math.u-bordeaux1.fr}}
\end{center}

\begin{abstract}
  We extend Penrose's peeling model for the asymptotic behaviour of
  solutions to the scalar wave equation at null infinity on
  asymptotically flat backgrounds, which is well understood for flat
  space-time, to Schwarzschild and the asymptotically simple
  space-times of Corvino-Schoen/Chrusciel-Delay.  We combine conformal
  techniques and vector field methods: a naive adaptation of the
  ``Morawetz vector field'' to a conformal rescaling of the
  Schwarzschild metric yields a complete scattering theory on
  Corvino-Schoen/Chrusciel-Delay space-times. A good classification of
  solutions that peel arises from the use of a null vector field that
  is transverse to null infinity to raise the regularity in the
  estimates. We obtain a new characterization of solutions admitting a
  peeling at a given order that is valid for both Schwarzschild and
  Minkowski space-times. On flat space-time, this allows larger
  classes of solutions than the characterizations used since Penrose's
  work. Our results establish the validity of the peeling model at all
  orders for the scalar wave equation on the Schwarzschild metric and
  on the corresponding Corvino-Schoen/Chrusciel-Delay space-times.
\end{abstract}

\tableofcontents

\section{Introduction}

Penrose's null infinity, $\scri$, of a Lorentzian space-time is a powerful tool for studying the asymptotics of massless fields in both flat and curved space-times. The asymptotic series of the physical field in the physical space-time translates, in an unphysical conformally rescaled space-time, into the Taylor series of the field off the finite hypersurface $\scri$. This can also be used as a basis for reformulating the scattering theory of massless fields into a Goursat problem based on $\scri$ (see Penrose 1963 \cite{Pe63}, Friedlander 1980 and 2001 \cite{Fri1980, Fri2001}, Baez, Segal and Zhou 1990 \cite{BaSeZo} and the authors 2004 \cite{MaN04}).  However, the use of $\scri$ in curved space-times has for a long time been controversial since firstly, it has not been clear that there is a good generic class of space-times with smooth null infinities and secondly, even if so, it is still not known whether  interesting massless fields do in fact fall off as proposed in the formalism so as to be amenable to analysis. The former problem has now been resolved in various ways, Christodoulou-Klainerman \cite{ChriKla}, Corvino \cite{Co2000}, Chrusciel and Delay \cite{ChruDe2002, ChruDe2003}, Corvino-Schoen \cite{CoScho2003}, Friedrich (see \cite{HFri2004} for a survey of his contributions) and Klainerman-Nicol\`o \cite{KlaNi,KlaNi2002,KlaNi2003}. As far as the second issue is concerned, schemes in which fields admit all kinds of singularities at null infinity have been proposed and shown to be consistent, at least in the neighbourhood of $\scri$. Such results have often been understood as an indication that the peeling model on non trivial asymptotically flat space-times may apply to more restrictive sets of data than in the flat case. To this day, even in the relatively simple case of the Schwarzschild metric, there is no precise understanding of the type of initial data that will guarantee the regularity of solutions on null infinity.

Minkowski space can be conformally embedded into the Einstein
cylinder, $\R \times S^3$ and compactified by adding $\scri^-$ (past null infinity), $\scri^+$ (future null infinity), which are
respectively the past and future light cones of a point at space-like
infinity (denoted $i^0$) in the conformal compactification and which refocus
to the vertices $i^-$ and $i^+$ (past and future timelike infinities). The points $i^\pm$ and
$i^0$ are smooth points of the conformal structure. In this picture,
it is easy to see that data for massless fields on a Cauchy surface
through $i^0$ will propagate smoothly up the Einstein cylinder
indefinitely if the data is smooth over the Cauchy surface including
$i^0$ in the conformally rescaled space-time. This shows that there
is a good class of massless fields that are smooth at $\scri$ in
Minkowski space, consisting of fields whose conformal rescaling admits
a smooth extension across $i^0$. Solutions that are regular on
Minkowski space but with arbitrarily bad behaviour at $\scri$ can be
understood in this picture as arising from data that is singular at
$i^0$ after conformal rescaling.

A non-trivial asymptotically simple curved space-time has almost the
same global structure with smooth conformal boundary $\scri^\pm$ and
smooth vertices $i^\pm$ but there is necessarily a singularity at
$i^0$. Black hole space-times such as Schwarzschild or Kerr have a
more complicated conformal geometry, with singular $i^\pm$, but in the
neighbourhood of $i^0$, they are similar to asymptotically simple
space-times. The singularity at $i^0$ is associated with the ADM mass
and so cannot be removed in a physical context except in the case of
flat space-time. Thus the above conformal argument for the existence
of a large class of solutions that are smooth at $\scri$ needs to be
modified~: using naive methods, one can only guarantee that solutions
are smooth on $\scri$ if they are compactly supported away from $i^0$.
The general understanding in such situations is that if solutions only
fall off at some finite rate near $i^0$ on spacelike slices, the rate
of fall-off at spacelike infinity will determine the regularity of the
solution at $\scri$. However a precise quantitative description of
this relation has not previously been available and very different
opinions have been expressed as to what it should be.

This paper provides this description completely for scalar fields on
Schwarzschild space-time. This immediately extends to the
corresponding Corvino-Schoen/Chrusciel-Delay space-times
(asymptotically simple space-times that are diffeomorphic to
Schwarzschild outside a compact set). The essential tool is energy
estimates, also referred to as vector field methods, with the
additional feature that we apply these techniques on a conformally
rescaled space-time and not on the physical space-time. We choose
carefuly a vector field to contract with the stress-enery tensor~: the
associated energy must be positive on spacelike slices and on $\scri$,
and the asymptotic behaviour of the vector field near $\scri$ and
$i^0$ has to be chosen so as to keep a uniform control on the
estimates in these regions. The ``Morawetz vector field'', introduced
by Morawetz in the early 1960's (see \cite{Mo1962}), is a well-known
tool for obtaining pointwise decay estimates in flat space-time. It
can be easily adapted to the Schwarzschild case and has indeed already
been used in this setting to obtain pointwise decay rates for
spherically symmetric equations by Dafermos and Rodnianski
\cite{DaRo}. The Schwarzschild version of the Morawetz vector field
allows us to obtain basic energy estimates between $\scri$ and some
initial data surface. Higher order estimates are then deduced by
commuting into the equation a null vector field transverse to $\scri$,
and not the Morawetz vecor field itself, contrary to what can be done
for flat space-time. This provides a rigorous definition of the
peeling as well as the precise function spaces of initial data giving
rise to solutions that peel at any given order. We show that these
function spaces are optimal for our definition. Comparing them to the
usual peeling picture on Minkowski space-time, it turns out that our
classes of data giving rise to a given regularity of the solution on
$\scri$ are larger than what was previously known for flat space-time.
Our results therefore validate the peeling model for the wave equation
on the Schwarzschild metric at all orders~: they also provide a new
definition of the peeling that is more precise than the definitions
used sofar. A remarkable feature of this new definition is that, for two different values of the mass $m$ of the black hole, the spaces of initial data for a given regularity are canonically isomorphic. Moreover, the equivalence in the norms is uniform on any compact interval in $m$, typically $m \in [0,M]$, $M>0$, which includes Minkowski space.

The paper is organized as follows. Section \ref{MinkPeel} describes the peeling on Minkowski space as it is usually understood, then gives an alternative description in terms of vector field methods. Section \ref{BasicFormulae} contains the basic ingredients that will be used in the rest of the paper~: a foliation of a neighbourhood of $i^0$ in Schwarzschild's space-time is given, a careful choice of identifying vector field is made for this foliation (this will be crucial for higher order estimates), the Morawetz vector field is introduced and explicit formulae are obtained for energy densities and error terms. The fundamental energy estimates are derived in section \ref{FundEst} and it is remarked that they entail a conformal scattering theory for the wave equation on Corvino-Schoen/Chrusciel-Delay space-times. Higher order estimates are then obtained in section \ref{Peeling}, giving a complete classification of the spaces of data that give rise to solutions that peel at any order~; these spaces are expressed in definition \ref{PeelingSpaces} that concludes the section. The results are interpreted in section \ref{Interpretation}. The appendix contains the proofs of the theorems.

\noindent {\bf Important remarks.}

\noindent $\bullet$ All our results are focused on a neighbourhood of spacelike infinity $i^0$, where the difficulty is localized. Our definition of the peeling is therefore concerned with the regularity on null infinity near $i^0$. Global regularity on $\scri$ can then easily be recovered by assuming in addition that the data are in local Sobolev spaces on a Cauchy hypersurface.

\noindent $\bullet$ We work with the wave equation in this paper since some pathological schemes have been put forward in this case. We would expect the situation to be simpler for higher helicity massless fields. This will be the subject of a subsequent study.

\noindent $\bullet$ Throughout the paper, the energy estimates are performed for solutions associated with smooth compactly supported data~; a completion of the space of smooth compactly supported functions in the norms given by these estimates then give the function spaces that characterize data giving rise to solutions that peel at a given order.

\noindent $\bullet$ For more details on the conformal compactifications of the Schwarzschild and Min\-kow\-ski space-times that we use, see Penrose and Rindler \cite{PeRi84}.

\noindent {\bf Notations.}

\noindent $\bullet$ We shall use the notation $\lesssim$ to signify that the left hand side is bounded by a positive constant times the right hand side, the constant being independent of the parameters and functions appearing in the inequality.

\noindent $\bullet$ Given $\cal M$ a smooth manifold and $\cal O$ an open set of $\cal M$, we denote by ${\cal C}^\infty_0 ({\cal O})$ the space of smooth functions with compact support on ${\cal O}$, and by ${\cal D}' ({\cal O})$ its topological dual, the space of distributions on ${\cal O}$.

\section{Peeling on flat space-time} \label{MinkPeel}

We denote by $\mathbb{M}$ Minkowski space and by $\eta$ the Minkowski metric. The peeling properties of solutions to field equations on Minkowski space can be easily understood as a consequence of the embedding of $\mathbb{M}$ into the Einstein cylinder. We first recall the conformal compactification of Minkowski space that realizes this embedding, then we proceed with the usual description of the peeling in this framework. Finally, we propose an alternative description based on vector field methods.

\subsection{Conformal compactification of Minkowski space}

We choose the advanced and retarded coordinates
\[ u= t-r \, ,~ v=t+r \, ,\]
then we put
\begin{gather*}
p = \arctan u \, ,~ q = \arctan v \, , \\
\tau = p+q = \arctan (t-r) + \arctan (t+r ) \, ,~ \\ \zeta = q-p = \arctan (t+r ) - \arctan (t-r) \, .
\end{gather*}
Choosing the conformal factor
\begin{equation} \label{FullConfFact}
\Omega = \frac{2}{\sqrt{1+u^2} \sqrt{1+v^2}} \, ,
\end{equation}
we obtain the rescaled metric
\begin{equation}
\mathfrak{e} = \Omega^2 \eta = \d \tau^2 - \d \zeta^2 - \frac{(v-u)^2}{(1+u^2)(1+v^2)} \, \d \omega^2 = \d \tau^2 - \d \zeta^2 - \sin^2 \zeta \, \d \omega^2
\end{equation}
and Minkowski space is now described by the domain
\[ \mathbb{M} = \{ |\tau| + \zeta \leq \pi \, ,~ \zeta \geq 0 \, ,~ \omega \in S^2 \} \, . \]
The metric $\mathfrak{e}$ is the Einstein metric $\d \tau^2 -\sigma^2_{S^3}$, where $\sigma^2_{S^3}$ is the Euclidian metric on the $3$-sphere, and it extends analytically to the whole Einstein cylinder $\mathfrak{E} = \R_\tau \times S^3_{\zeta, \theta , \varphi}$. The full conformal boundary of Minkowski space can be defined in this framework. Several parts can be distinguished.
\begin{itemize}
\item Future and past null infinities~:
\begin{eqnarray*}
\scri^+ &=& \left\{ (\tau \, ,~ \zeta \, ,~ \omega ) \, ;~ \tau + \zeta = \pi \, ,~ \zeta \in ] 0,\pi [ \, ,~ \omega \in S^2 \right\} \, ,\\
\scri^- &=& \left\{ (\tau \, ,~ \zeta \, ,~ \omega ) \, ;~ \zeta -\tau = \pi \, ,~ \zeta \in ] 0,\pi [ \, ,~ \omega \in S^2 \right\} \, .
\end{eqnarray*}
They are smooth null hypersurfaces for $\mathfrak{e}$.
\item Future and past timelike infinities~:
\[ i^\pm = \left\{ (\tau =\pm \pi \, ,~ \zeta =0 \, ,~ \omega ) \, ;~ \omega \in S^2 \right\} \, . \]
They are smooth points for $\mathfrak{e}$.
\item Spacelike infinity~:
\[ i^0 = \left\{ (\tau =0 \, ,~ \zeta =\pi \, ,~ \omega ) \, ;~ \omega \in S^2 \right\} \, . \]
It is also a smooth point for $\mathfrak{e}$.
\end{itemize}
The hypersurface $\{ t=0\}$ in Minkowski space is described by the $3$-sphere $\{ \tau =0 \}$ minus the point $i^0$ on the Einstein cylinder.

The scalar curvature of $\mathfrak{e}$ can be calculated easily using the conformal factor $\Omega$~:
\begin{equation} \label{ScalEinstein}
\frac{1}{6} \mathrm{Scal}_\mathfrak{e} = \Omega^{-3} \square_\eta \Omega = 1 \, .
\end{equation}
In this framework, the vector field $\partial_\tau$ is Killing since the metric $\mathfrak{e}$ does not depend on $\tau$.

\subsection{Conformal rescaling of the wave equation}

The conformal invariance of the wave equation entails the equivalence of the two properties~:
\begin{description}
\item[(i)] $\tilde{\psi} \in {\cal D}' (\mathbb{M} )$ satisfies
\begin{equation} \label{MinkWaveEq}
\square_\eta \tilde{\psi} = 0 \, ,~ \square_\eta = \partial_t^2 - \frac{1}{r^2} \partial_r r^2 \partial_r - \frac{1}{r^2} \Delta_{S^2} \, ;
\end{equation}
\item[(ii)] $\psi := \Omega^{-1} \tilde{\psi}$ satisfies
\begin{equation} \label{EinstWaveEq}
\square_\mathfrak{e} \psi + \psi = 0 \, ,~ \square_\mathfrak{e} = \partial_\tau^2 - \Delta_{S^3} \, .
\end{equation}
\end{description}

\subsection{The usual description of peeling}

The observation of the peeling in Minkowski space is usually derived from the property that the Cauchy problem on the Einstein cylinder is well-posed in ${\cal C}^k$ spaces. This are completely standard result, we just quote it here.
\begin{proposition} \label{CPCk}
Let $k \in \N^*$. For any initial data $\psi_0 \in {\cal C}^k (S^3)$, $\psi_1 \in {\cal C}^{k-1} (S^3)$, there exists a unique solution $\psi \in {\cal C}^k (\mathfrak{E})$ of (\ref{EinstWaveEq}) such that $\psi (0) = \psi_0$ and $\partial_\tau \psi (0) = \psi_1$.
\end{proposition}
This provides a natural definition of solutions that peel at a given order $k \in \N$~:
\begin{definition} \label{DefPeelMinkCk}
A solution $\tilde{\psi}$ of (\ref{MinkWaveEq}) is said to peel at order $k \in \N$ if $\psi =\Omega^{-1} \tilde{\psi}$ extends as a ${\cal C}^k$ function on the whole Einstein cylinder. The latter property is satisfied by solutions $\psi$ of (\ref{EinstWaveEq}) arising from initial data $\psi |_{\tau=0} \in {\cal C}^k (S^3)$ and $\partial_\tau \psi |_{\tau=0} \in {\cal C}^{k-1} (S^3)$. Going back to Minkowski space and to the physical field $\tilde{\psi}$, this gives us a corresponding class of data for (\ref{MinkWaveEq}), giving rise to solutions that peel at order $k$.
\end{definition}

\subsection{Description by means of vector field methods}

Although it is not commonly used, we can give an analogous description of the peeling in Minkowski space in terms of Sobolev spaces instead of ${\cal C}^k$ spaces. To do so, we write equation (\ref{EinstWaveEq}) in its hamiltonian form
\[ \frac{\partial}{\partial \tau} \left( \begin{array}{c} \psi \\ \partial_\tau \psi \end{array} \right) = iH \left( \begin{array}{c} \psi \\ \partial_\tau \psi \end{array} \right) \, ,~ H = -i \left( \begin{array}{cc} 0 & 1 \\ {\Delta_{S^3} -1} & 0 \end{array} \right) \]
and work on the Hilbert space
\[ {\cal H} = H^1 (S^3 ) \times L^2 (S^3 ) \]
with the usual inner product
\[ \left< \left( \begin{array}{c} f_1 \\ f_2 \end{array} \right) , \left( \begin{array}{c} g_1 \\ g_2 \end{array} \right) \right>_{\cal H} = \int_{S^3} \left( \nabla_{S^3} f_1 . \nabla_{S^3} \bar{g_1} + f_1 \bar{g_1} + f_2 \bar{g_2} \right) \d \mu_{S^3} \, ,\]
where $\nabla_{S^3}$ is the Levi-Civitta connection and $\mu_{S^3}$ the measure induced by the Euclidian metric on $S^3$. We have~:
\begin{proposition} \label{CPHk}
The operator $H$ with its natural domain $D(H) = H^2 (S^3 ) \times H^1 (S^3 )$ is self-adjoint on ${\cal H}$ and its successive domains are
\[ D(H^k) = H^{k+1} (S^3 ) \times H^k (S^3 ) \, .\]
Let $k \in \N^*$. For any initial data $\psi_0 \in H^k (S^3)$, $\psi_1 \in H^{k-1} (S^3)$, there exists a unique solution
\[ \psi \in \bigcap_{l=0}^k {\cal C}^l \left( \R_\tau \, ;~ H^{k-l} (S^3) \right) \]
of (\ref{EinstWaveEq}) such that $\psi (0) = \psi_0$ and $\partial_\tau \psi (0) = \psi_1$. In particular, $\psi \in H^k_\mathrm{loc} (\mathfrak{E})$. Moreover, for any $0\leq l \leq k-1$, $\| \psi (\tau ) \|^2_{H^{l+1} (S^3)} + \| \partial_\tau \psi (\tau ) \|^2_{H^{l} (S^3)}$ is constant throughout time.
\end{proposition}
This gives us a definition of solutions that peel at a given order $k \in \N$ that is analogous to \ref{DefPeelMinkCk}, but now in terms of Sobolev spaces~:
\begin{definition} \label{DefPeelMinkHk}
A solution $\tilde{\psi}$ of (\ref{MinkWaveEq}) is said to peel at order $k \in \N$ if $\psi =\Omega^{-1} \tilde{\psi}$ extends as a function that is in $H^{k+1}_\mathrm{loc}$ on the whole Einstein cylinder. The latter property is satisfied by solutions $\psi$ of (\ref{EinstWaveEq}) arising from initial data $\psi |_{\tau=0} \in H^{k+1} (S^3)$ and $\partial_\tau \psi |_{\tau=0} \in H^{k} (S^3)$. Going back to Minkowski space and to the physical field $\tilde{\psi}$, this gives us a corresponding class of data for (\ref{MinkWaveEq}), giving rise to solutions that peel at order $k$.
\end{definition}
What is essentially unsatisfactory in definitions \ref{DefPeelMinkCk} and \ref{DefPeelMinkHk} is that they merely provide an inclusion~: we know a class of data that gives rise to peeling at order $k$, but we do not know whether we have all such data. An alternative approach consists in using vector field methods (energy estimates). Such techniques allow to prove easily the last property in proposition \ref{CPHk} but are much more flexible than a purely spectral result~: we can just as naturally obtain estimates between the initial data surface and $\scri^+$. This will provide us with a third description of the peeling on flat space-time. It will be more precise than the first two in that the set of suitable data for a peeling at order $k$ will be completely characterized, and not just a subset of it.

We consider the stress energy tensor for equation (\ref{EinstWaveEq})
\begin{equation} \label{SETensorEinsWE}
T_{ab} = T_{(ab)} = \partial_a \psi \partial_b \psi - \frac12 \mathfrak{e}_{ab} \mathfrak{e}^{cd} \partial_c \psi \partial_d \psi +\frac{1}{2} \psi^2 \mathfrak{e}_{ab}
\end{equation}
and contract it with the Killing vector field $\partial_\tau$. This yields the conservation law
\begin{equation} \label{EinsteinConsLaw}
\nabla^a \left( K^b T_{ab} \right) = 0 \, .
\end{equation}
The energy $3$-form $K^aT_{ab} \d^3 x^b = K^a T_a^b \partial_b \hook \mathrm{dVol}^4$ has the expression
\begin{equation} \label{EinsteinEn3Form}
K^aT_{ab} \d^3 x^b = \psi_\tau \nabla \psi \hook \mathrm{dVol}^4 + \frac{1}{2} \left( - \psi_\tau^2 + \left| \nabla_{S^3} \psi \right|^2 + \psi^2 \right) \partial_\tau \hook \mathrm{dVol}^4 \, .
\end{equation}
Integrating (\ref{EinsteinEn3Form}) on an oriented hypersurface $S$ defines the energy flux across this surface, denoted ${\cal E}_S (\psi )$. For instance, denoting $X_\tau = \{ \tau \} \times S^3$ the level hypersurfaces of the function $\tau$ 
\[ {\cal E}_{X_\tau} (\psi ) = \frac{1}{2} \int_{X_\tau} \left( \psi_\tau^2 + \left| \nabla_{S^3} \psi \right|^2 + \psi^2 \right) \d \mu_{S^3} \, ,\]
and parametrizing $\scri^+$ as $\tau = \pi - \zeta$,
\begin{eqnarray*}
{\cal E}_{\scri^+} (\psi ) &=& \frac{1}{\sqrt{2}} \int_{\scri^+} \left( -2 \psi_\tau \psi_\zeta + \psi_\tau^2 + \left| \nabla_{S^3} \psi \right|^2 + \psi^2 \right) \d \mu_{S^3} \\
&=& \frac{1}{\sqrt{2}} \int_{\scri^+} \left( \left| \psi_\tau - \psi_\zeta \right|^2 + \frac{1}{\sin^2 \zeta} \left| \nabla_{S^2} \psi \right|^2 + \psi^2 \right) \d \mu_{S^3} \, .
\end{eqnarray*}
This is a natural $H^1$ norm of $\psi$ on $\scri^+$, involving only the tangential derivatives of $\psi$ along $\scri^+$.

Now consider a smooth solution $\psi$ of (\ref{EinstWaveEq}). The conservation law (\ref{EinsteinConsLaw}) tells us that (\ref{EinsteinEn3Form}) is closed, hence, integrating it on the closed hypersurface made of the union of $X_0$ and $\scri^+$, we obtain
\[ {\cal E}_{\scri^+} (\psi ) = {\cal E}_{X_0} (\psi ) \]
and since $\partial_\tau$ is a Killing vector, for any $k\in \N$, $\partial_\tau^k \psi$ satisfies equation (\ref{EinstWaveEq}), whence
\[ {\cal E}_{\scri^+} (\partial_\tau^k \psi ) = {\cal E}_{X_0} (\partial_\tau^k \psi ) \, . \]
Using equation (\ref{EinstWaveEq}), for $k= 2p$, $p\in \N$, we have
\begin{eqnarray}
{\cal E}_{X_0} (\partial_\tau^k \psi ) &=& \| \partial_\tau^{2p} \psi \|_{H^1 (X_0 )}^2 + \| \partial_\tau^{2p+1} \psi \|_{L^2 (X_0 )}^2 \nonumber \\
&=&  \| (1-\Delta_{S^3} )^{p} \psi \|_{H^1 (X_0)}^2 + \| (1-\Delta_{S^3} )^{p} \partial_\tau \psi \|_{L^2 (X_0)}^2 \nonumber \\
&=& \| \psi \|_{H^{2p+1} (X_0)}^2 + \| \partial_\tau \psi \|_{H^{2p} (X_0)}^2 \, , \label{EinsteinEnEqEven}
\end{eqnarray}
and for $k=2p+1$, $p \in \N$,
\begin{eqnarray}
{\cal E}_{X_0} (\partial_\tau^k \psi ) &=& \| \partial_\tau^{2p+1} \psi \|_{H^1 (X_0 )}^2 + \| \partial_\tau^{2p+2} \psi \|_{L^2 (X_0 )}^2 \nonumber \\
&=& \| (1-\Delta_{S^3} )^{p} \partial_\tau \psi \|_{H^1 (X_0)}^2 + \| (1-\Delta_{S^3} )^{p+1} \psi \|_{L^2 (X_0)}^2 \nonumber \\
&=& \| \psi \|_{H^{2p+2} (X_0)}^2 + \| \partial_\tau \psi \|_{H^{2p+1} (X_0)}^2\, . \label{EinsteinEnEqOdd}
\end{eqnarray}
Hence, we have for each $k \in \N$~:
\begin{equation} \label{EinsteinEnEq}
\| \psi \|^2_{H^{k+1} (X_0 )} + \| \partial_\tau \psi \|^2_{H^{k} (X_0 )} \simeq  {\cal E}_{X_0} (\partial_\tau^k \psi ) = {\cal E}_{\scri^+} (\partial_\tau^k \psi ) \simeq \| \partial^k_\tau \psi \|^2_{H^{1} (\scri^+ )} \, .
\end{equation}
So the transverse regularity on $\scri^+$ (here described by ${\cal E}_{\scri^+} (\partial_\tau^k \psi )$), is entirely determined by the Sobolev spaces in which the initial data lie. We use this equivalence to formulate our third definition of peeling at order $k$ and to describe the exact class of data that gives rise to this property.

\begin{definition} \label{DefPeelMinkEnEst}
A solution $\tilde{\psi}$ of (\ref{MinkWaveEq}) is said to peel at order $k \in \N$ if the trace on $\scri^+$ of $\partial_\tau^k \psi$ is in $H^{1} (\scri^+)$. The set of solutions of (\ref{EinstWaveEq}) satisfying this property is exactly the set of solutions whose data at $\tau =0$ satisfy $\psi |_{\tau=0} \in H^{k+1} (S^3)$ and $\partial_\tau \psi |_{\tau=0} \in H^{k} (S^3)$. Going back to Minkowski space and to the physical field $\tilde{\psi}$, this gives us the exact class of data for (\ref{MinkWaveEq}), giving rise to solutions that peel at order $k$.
\end{definition}
\begin{remark} \label{Mink4VolPeeling}
The description given in definition \ref{DefPeelMinkHk} corresponds to the slightly weaker approach, via the equality
\[ {\cal E}_{X_\tau} (\partial_\tau^k \psi )= {\cal E}_{X_0} (\partial_\tau^k \psi )~\forall~ \tau \in \R \, ,\]
which entails
\begin{equation} \label{4DimEstimate}
\left\| \psi \right\|_{H^{k+1} (\Omega^+ )} \lesssim \| \psi \|^2_{H^{k+1} (X_0 )} + \| \partial_\tau \psi \|^2_{H^{k} (X_0 )} \, ,
\end{equation}
where $\Omega^+$ is the $4$-volume in the future of $X_0$ and the past of $\scri^+$. It is slightly weaker in the way we understand the transverse regularity at $\scri^+$ (implicitly in terms of trace theorems for Sobolev spaces), hence the fact that we have merely inequalities instead of equivalences. But the spaces of initial data for which regularity at a given order is guaranteed near $\scri^+$ are the same.
\end{remark}

\section{Basic formulae} \label{BasicFormulae}

The Schwarzschild metric is
$$
g=\left( 1-\frac{2m}r \right) \d t^2- \left( 1-\frac{2m}r\right)^{-1} \d r^2 -r^2\d\omega^2
$$
where $\d\omega^2$ is the Euclidian metric on the unit sphere $S^2$.  The associated d'Alembertian is
\[ \square_g = \left( 1 -\frac{2m}{r} \right)^{-1} \frac{\partial^2}{\partial t^2} - \frac{1}{r^2} \frac{\partial}{\partial r} \, r^2 \left( 1 -\frac{2m}{r} \right) \frac{\partial}{\partial r} - \frac{1}{r^2} \Delta_{S^2} \, ,\]
where $\Delta_{S^2}$ is the Laplacian on $S^2$ endowed with the Euclidian metric. Putting
$$
R=1/r \, ,~ u=t-r_* \, ,~\mathrm{with~} r_*=r+2m\log(r-2m)\, ,
$$
the metric can be transformed and conformally rescaled to give 
$$
\hat{g}= R^2 g= R^2(1-2mR)\d u^2-2\d u\d R-\d\omega^2
$$
and in this form of the metric, $\scri^+$ is given by $R=0$. We denote by $G$ the function that to $r$ associates $r_*$~:
\begin{equation} \label{FunctionRstar}
G\, :~]2m +\infty [ \longrightarrow \R \, ,~ G(r) = r+2m\log(r-2m) \, ;
\end{equation}
it is an analytic diffeomorphism between $]2m , +\infty [$ and $\R$. The scalar curvature of $\hat{g}$ is given by
\[ \frac{1}{6} \mathrm{Scal}_{\hat{g}} = R^{-3} \square_g R =2mR \, .\]
The inverse metric is~:
$$
\hat g^{-1} =-2\p_u\p_R-R^2(1-2mR)\p_R^2-\eth\bar\eth \, ,
$$
which gives the d'Alembertian
\[ \square_{\hat{g}} = -2 \partial_u \partial_R - \partial_R R^2 \left( 1-2mR
\right) \partial_R - \Delta_{S^2} \, .\]
The equation that we study is the conformally invariant wave equation
\begin{equation} \label{ConfWaveEq}
\left( \square_{\hat{g}} + \frac{1}{6} \mathrm{Scal}_{\hat{g}} \right) \phi = \left( \square_{\hat{g}} + 2mR \right) \phi = 0 \, .
\end{equation}
It is such that, denoting by ${\cal B}_I$ (for block $I$) the exterior of the black hole in Schwarzschild's space-time, the two following properties are equivalent~:
\begin{enumerate}
\item $\tilde{\phi} \in {\cal D}' ({\cal B}_I )$ satisfies $\square_g \tilde{\phi} =0$\, ;
\item $\phi := R^{-1} \tilde{\phi} = r\tilde{\phi}$ satisfies (\ref{ConfWaveEq}) on ${\cal B}_I$.
\end{enumerate}

\subsection{Stress-energy tensor and conservation laws}

We use the stress energy tensor for the free wave equation $\square_{\hat{g}} \phi =0$~:
\begin{equation} \label{StressEnTensor}
T_{ab}=T_{(ab)}=\partial_a \phi \, \partial_b \phi - \frac12 \hat{g}_{ab} \hat{g}^{cd} \partial_c \phi \, \partial_d \phi \, ,
\end{equation}
(although we could equivalently use the one for the Klein-Gordon equation $\square_{\hat{g}} \phi + \phi =0$, obtained by adding to $T_{ab}$ the term $\frac{1}{2} \phi^2 \hat{g}_{ab}$). It satisfies
\[ \nabla^a T_{ab} = \square_{\hat{g}} \phi \partial_b \phi = - 2mR \phi \partial_b \phi\, .\]
This gives rise to approximately conserved quantities for our equation by considering the integral of the energy $3$-form
\begin{equation} \label{En3Form}
K^aT_{ab}\d^3x^b=K^aT_a^b\p_b\hook \d \mathrm{vol}^4 \, ,
\end{equation}
where $K^a$ is a Killing or approximate Killing vector. We can calculate the error in the conservation law as follows,
\begin{equation} \label{ApproxConsLaw}
 \nabla^a \left( K^b T_{ab} \right) = \square_{\hat{g}} \phi K^b\partial_b \phi + \nabla^a K^b T_{ab} = -2mR \phi K^b\partial_b \phi + (\nabla_aK_b)T^{ab} \, .
\end{equation}
If $K^a$ is Killing, the Killing form, also referred to as deformation tensor, $\nabla_{(a} K_{b)}$, vanishes and so does $(\nabla_aK_b)T^{ab}$, but this still does not give us an exact conservation law because of the term involving the scalar curvature.

We shall use this to perform energy estimates in a neighbourhood of $i^0$~: for a choice of foliation, we estimate the error term by the energy on each slice and then invoke Gronwall's lemma. We also, however, need to be clear about the choice of vector field used to identify the different slices.

\subsection{Foliation and identifying vector field}

Our essential foliation is by the spacelike hypersurfaces
\begin{equation} \label{FoliationH}
{\cal H}_s = \left\{ u=-sr_* \right\} \, , ~\mathrm{with~} {\cal H}_0 =\scri^+ ~\mathrm{and~} {\cal H}_1 = \{ t=0 \} \, .
\end{equation}
We record the relations
\[\frac{\d r_*}{\d R}=\frac{-1}{R^2(1-2mR)} \, ,~ \left. \d R \right|_{{\cal H}_s} = \frac{R^2 (1-2mR)}{s} \left. \d u \right|_{{\cal H}_s}= \frac{r_* R^2 (1-2mR)}{|u|} \left. \d u \right|_{{\cal H}_s} ~\mathrm{near~} i^0 \, .\]
This foliation is not smooth in the sense that $r_*^{-1}$ is not a smooth function of $R$ at $R=0$~; although the first derivative is bounded and tends to 1, the second is logarithmically divergent as $R\rightarrow 0$.

In this case, a natural identifying vector field $\nu$ needs to satisfy $\nu (s) =1$~; we choose
\begin{equation} \label{VectVHs}
\nu=r_*^2R^2(1-2mR)|u|^{-1}\p_R \, .
\end{equation}
The $4$-volume measure $\mathrm{dVol}^4$, against which the error terms will be integrated, is thus decomposed into the product of $\d s$ and $\nu \hook \mathrm{dVol}^4 |_{{\cal H}_s}= r_*^2 R^2 (1-2mR) |u|^{-1} \d u \d^2\omega |_{{\cal H}_s}$, the former being the measure along the integral lines of $\nu^a$ and the latter our $3$-volume measure on each ${\cal H}_s$.

Such a choice of identifying vector field $\nu^a$, which is really associated with the choice of parameter $s$ for the foliation, will lead to error terms that cannot be controlled by the energy density\footnote{To be more precise, this does not occur for the fundamental estimates, because the scalar curvature $2mR$ gives us some extra fall-off at $\scri$. For higher order estimates, commuting $\partial_R$ into the equation will give error terms without any fall-off. So the problem will occur as soon as we try to gain one extra degree of regularity from the fundamental estimates.} and therefore to the impossibility of performing a priori estimates. All we need to do in order to solve this problem is the following change of parameter~:
\begin{equation} \label{ParamTau}
\tau := -2\left( \sqrt{s} -1 \right) \, ,~ \mbox{so that } \frac{\d \tau}{\d s} = -\frac{1}{\sqrt{s}}.
\end{equation}
The change of sign and the $-1$ term are there purely for aesthetic reasons, the important part is $2\sqrt{s}$. This new parameter varies from $0$ to $2$ as $s$ varies from $1$ to $0$. We denote
\begin{equation} \label{FoliationSigma}
\Sigma_{\tau (s)} = {\cal H}_s \, .
\end{equation}
The natural new identifying vector field is
\begin{equation} \label{VectVSigmaTau}
V = -\sqrt{s}\, \nu = -\sqrt{\frac{|u|}{r_*}} r_*^2 R^2(1-2mR)|u|^{-1}\p_R = -\left( r_* R \right)^{3/2} (1-2mR) \sqrt{\frac{R}{|u|}} \, \p_R \, .
\end{equation}
We shall use both notations $\{ {\cal H}_s \}_s$ and $\{ \Sigma_\tau \}_\tau$, i.e. both parameters $s$ and $\tau$ related by (\ref{ParamTau}), for our foliation by spacelike hypersurfaces. The parameter $s$ has a straightforward definition, it is useful for calculations on a single hypersurface and for the sake of simplicity we systematically use it for expressing the main results. The parameter $\tau$ on the other hand is adapted to a priori estimates and will be used in the proof of energy estimates.

As we shall work on a small neighbourhood of $i^0$ of the form
\[ \Omega_{u_0}^+ := \left\{ u<u_0 \, , ~ t>0 \right\} \, , \]
with $u_0 <<-1$, we need to consider the hypersurface
\[ {\cal S}_{u} = \left\{ (u ,R,\omega ) \, ;~ 0 <R< 1/ G^{-1} (-u) \,
  ,~ \omega \in S^2 \right\} \, ,\] 
that forms, for $u=u_0$, the part of the null boundary of $\Omega_{u_0}^+$
in the finite Schwarzschild space-time ($G$
is the function (\ref{FunctionRstar}) defining $r_*$ in terms of
$r$). The natural vector field connecting one surface of constant $u$
to the next is 
\begin{equation} \label{VectWSu}
W:=\p_u+\frac{R^2(1-2mR)}{s} \p_R = \p_u + \frac{(r_*R)^2 (1-2mR)}{|u|r_*} \p_R\, , \mbox{ where } s=-\frac{u}{r_*} \, .
\end{equation}
The vector field $W$ is tangent to each hypersurface ${\cal H}_s$ and on $\scri^+$ reduces to $\partial_u$. The corresponding decomposition of the $4$-volume measure will be the product of $\d u$ along the integral lines of $W$ and of the $3$-volume measure on each ${\cal S}_u$~:
\[ \left. W \hook \mathrm{dVol}^4 \right|_{{\cal S}_u} = \left. \left( \d R \, \d^2\omega - \frac{ R^2(1-2mR)}{s} \, \d u \d^2 \omega \right) \right|_{{\cal S}_u} = \left. \d R \, \d^2\omega \right|_{{\cal S}_u}\, .\]

\subsection{The Morawetz vector field}

For $m=0$, $g$ is the Minkowski metric and the Morawetz vector field (see \cite{Mo1962})
\begin{eqnarray*}
K^a \partial_a &=& u^2 \partial_u + v^2 \partial_v \, ,~ u= t-r \, ,~ v=t+r \, , \\
&=& (r^2+t^2) \partial_t + 2tr \partial_r \, ,
\end{eqnarray*}
is a Killing vector for the rescaled metric $\hat{g}$. Its expression in terms of the variables $u=t-r$ and $R=1/r$ is
\begin{equation} \label{Morawetz}
K^a \partial_a = u^2 \partial_u - 2(1+uR) \partial_R \, .
\end{equation}
It is a classic vector field for pointwise decay estimates in flat space-time and a version of $K^a$ on the Schwarzschild space-time has been used recently for pointwise decay estimates on the Schwarzschild metric by Dafermos-Rodnianski \cite{DaRo}. One of its key features is that it is transverse to $\scri^+$, it will therefore give us more information on $\scri^+$ than other Killing vector fields such as for instance $\partial_u = \partial_t$.
\begin{remark}
 Expressing the time translation on the Einstein cylinder $\partial_\tau$ in the coordinate basis $u,R,\omega$, we obtain
\[ \partial_\tau = \frac{1}{2} \left[ \partial_u + u^2 \partial_u -2 (1+uR) \partial_R \right] \, . \]
We see that
\[ 2 \partial_\tau = \partial_u + K^a \partial_a \, , \]
or in terms of variables $t,r,\omega$,
\[ 2 \partial_\tau = \partial_t + K^a \partial_a \, ,\]
where $K$ is the Morawetz vector field (\ref{Morawetz}). So the two
vectors $\partial_\tau$ and $K^a$ are very close. We could have chosen
to use $\partial_\tau$ instead of $K^a$, the results and their proofs
would have been essentially unchanged. 
\end{remark}

We choose our approximate Killing vector field on the Schwarzschild metric as follows~: we keep the expression (\ref{Morawetz}) of the Morawetz vector field, but now in terms of the variables $u=t-r_*$ and $R=1/r$~; we put
\begin{equation} \label{VectorT}
T^a \partial_a :=u^2\p_u -2(1+uR)\p_R \, .
\end{equation}
Expressed in the coordinate system $(t,r_*)$, this gives
\begin{eqnarray*}
T^a \partial_a &=& (t-r_*)^2 \partial_t -2 \left( 1+uR \right)  \frac{-r^3}{r -2m} \left( \partial_t + \partial_{r_*} \right) \\
&=& (t^2+r_*^2)\partial_t -2tr_* \partial_t + 2 \left( 1-\frac{2m}{r} \right)^{-1} \left( tr +r^2 -rr_* \right) \left( \partial_t + \partial_{r_*} \right) \, .
\end{eqnarray*}
This is no longer a Killing vector field for $g$, its Killing form is
\begin{equation} \label{KillForm}
\nabla_{(a}T_{b)}= 4mR^2(3+uR)\d u^2 \, ,
\end{equation}
which has a good fall-off at infinity.

It is important to note that the vector field $T^a$ is uniformly timelike in a neighbourhood of $i^0$ and can therefore be used for obtaining energy estimates with positive definite energies on spacelike hypersurfaces.
\begin{remark} \label{TTimelike}
The vector field $T^a$ is everywhere timelike near $i^0$, indeed we have
\[ \hat{g}_{ab} T^a T^b = u^2 \left( 4 (1+uR) + u^2R^2 (1-2mR)\right) \, .\]
This vanishes for the two values of $uR$~:
\[ (uR)_\pm = -2 \frac{1\mp \sqrt{2mR}}{1-2mR}  \, , \]
that for small $R$ are arbitrarily close to $-2$, and $\hat{g}_{ab}
T^a T^b$ is positive for $uR$ outside $ [ (uR)_- , (uR)_+ ]$. In a
small enough neighbourhood of $i^0$, we shall have 
\[ -r_* < u << -1 \, ,~\mathrm{hence~} -\frac{r_*}{r} < Ru << -R \, ,\]
and $Ru$ therefore lives in an interval of the form $[-1-\varepsilon , 0]$, where $\varepsilon >0$ is as small as we wish it to be, since at infinity $r_* \simeq r$. Consequently, in a small enough neighbourhood of $i^0$, the vector $T^a$ is uniformly timelike.
\end{remark}

\subsection{Energy density and error terms for $K^a = T^a$}

The energy density 3-form $E (\phi )$ associated with $T^a$ is given by~:
\begin{eqnarray}
E (\phi )&:=&T^aT_{ab}\d^3x^b = T^aT_a^b\p_b\hook \d \mathrm{vol}^4\nonumber \\
&=&\left[ u^2\phi_u^2 +R^2(1-2mR)\left(u^2\phi_u\phi_R - (1+uR)
\phi_R^2\right) \right. \nonumber \\
&& \left. +(1+uR) |\nabla_{S^2}\phi|^2 \right]\d u \wedge \d ^2 \omega \nonumber \\
&&+\frac12\left[ \left((2+uR)^2 - 2mu^2R^3\right)\phi_R^2+u^2 |\nabla_{S^2}\phi|^2 \right]\d R \wedge \d^2 \omega \, . \label{En3FormT}
\end{eqnarray}
For a hypersurface $\cal S$, we denote
\[ {\cal E}_{\cal S} (\phi) := \int_{\cal S} E (\phi ) \, .\]
For instance,
\begin{gather}
{\cal E}_{\scri^+} (\phi ) = \int_{\scri^+} \left[ u^2\phi_u^2 + |\nabla_{S^2} \phi|^2 \right] \d u \wedge \d^2 \omega \, , \nonumber \\
{\cal E}_{{\cal S}_u} (\phi )= \int_{{\cal S}_u} \frac12\left[ \left((2+uR)^2 - 2mu^2R^3\right)\phi_R^2+u^2 |\nabla_{S^2}\phi|^2 \right]\d R \wedge \d ^2 \omega \, . \label{EnergySu}
\end{gather}
The energy on the surface ${\cal H}_s$ is given by
\begin{eqnarray}
{\cal E}_{{\cal H}_s} (\phi ) &=& \int_{{\cal H}_s} \left[ u^2 \phi_u^2 + R^2(1-2mR) u^2 \phi_u \phi_R \right. \nonumber \\
&& + R^2(1-2mR) \left( \frac{(2+uR)^2}{2s} -\frac{mu^2 R^3}{s} - (1+uR)
\right) \phi_R^2 \nonumber \\
&&  \left. + \left( \frac{u^2 R^2 (1-2mR)}{2s} + 1+uR \right) \left| \nabla_{S^2} \phi \right|^2 \right] \d u \wedge \d^2 \omega \, . \label{EnergyHs}
\end{eqnarray}
Using the expression (\ref{KillForm}) of the Killing form of $T^a$, the error term (\ref{ApproxConsLaw}) takes the form
\[ T_{ab}\nabla^aT^b -2mR \phi T^b\partial_b \phi = 4mR^2(3+uR)\phi_R^2 -2mR \phi \left( u^2 \partial_u \phi - 2 (1+uR) \partial_R \phi \right) \, . \]
Working on the foliation $\{ \Sigma_\tau \}_\tau$, this yields the error terms~:
\begin{eqnarray}
&&\left( T_{ab} \nabla^aT^b \right) V \hook \d\mathrm{vol}^4 = 4mR^2(3+uR) \phi_R^2 \left( r_* R \right)^{3/2} (1-2mR) \sqrt{\frac{R}{|u|}} \, \d u \wedge \d^2\omega \label{ErrTermNonKilling} \\
&&\left( -2mR \phi T^b\partial_b \phi \right) V \hook \mathrm{dVol}^4 = \nonumber \\
&& \hspace{0.5in} -2mR \phi \left( u^2 \partial_u \phi - 2 (1+uR) \partial_R \phi \right) \left( r_* R \right)^{3/2} (1-2mR) \sqrt{\frac{R}{|u|}} \, \d u \wedge \d^2\omega \, . \label{ErrTermOther}
\end{eqnarray}

\section{Fundamental estimates} \label{FundEst}

In this part, we obtain the basic energy estimates for equation (\ref{ConfWaveEq}). We work on the neighbourhood of $i^0$
\begin{eqnarray*}
\Omega^+_{u_0} &:=& \left\{ (u,R,\omega ) \, ;~ u<u_0 \, ,~ 0<s<1 \,
  ,~ \omega \in S^2 \right\}\, , \qquad s=-u/r^* \\
&=& \left\{ (u,R,\omega ) \, ;~ u<u_0 \, ,~ t>0 \, ,~ \omega \in S^2
\right\} \, , 
\end{eqnarray*}
for $u_0 <<-1$ such that $T^a$ is uniformly timelike in
$\Omega_{u_0}^+$. In order to prove the energy estimates, we use the
approximate conservation law (\ref{ApproxConsLaw})~; however, we must
first show that the energy on $\Sigma_\tau$ (\ref{EnergyHs}) uniformly
dominates the error term (\ref{ErrTermNonKilling}). On
$\Omega^+_{u_0}$ with $|u_0|$ large enough, this error term  is
uniformly equivalent to $R^2 \sqrt{R/|u|} \phi_R^2$, which is in turn
uniformly controlled by $R |u|^{-1} \phi_R^2$. The following lemma
shows that this is controlled by the energy density restricted to the
hypersurface ${\cal H}_s$. 
\begin{lemma} \label{LemmaNormeEquivEnergy}
On $\Omega_{u_0}^+$, for $u_0<0$, $| u_0|$ large enough, the energy density on ${\cal H}_s$ associated to $T^a$, i.e.
\[ \left. T^aT_{ab}\d^3x^b \right|_{{\cal H}_s \cap \Omega_{u_0}^+} \, ,\]
is uniformly equivalent to
\begin{equation} \label{NormeEquivEnergy}
\left. \left\{ \left( u^2\phi_u^2 + \frac{R}{|u|} \phi_R^2+ \left| \nabla_{S^2} \phi \right|^2 \right)  \d u\wedge\d^2\omega \right\} \right|_{{\cal H}_s \cap \Omega_{u_0}^+}  \, .
\end{equation}
\end{lemma}
For the error term (\ref{ErrTermOther}) involving the scalar curvature, the factor
\[ \left( T^b\partial_b \phi \right) V \hook \mathrm{dVol}^4 = \left( u^2 \partial_u \phi - 2 (1+uR) \partial_R \phi \right) \left( r_* R \right)^{3/2} (1-2mR) \sqrt{\frac{R}{|u|}} \, \d u \d^2\omega \]
is naturally controlled by the energy, thanks to the choice of parameter $\tau$, which gives us the factor $\sqrt{R/|u|}$. Indeed its two terms are respectively equivalent to $u^2 \sqrt{R/|u|} \phi_u$, which is unifomly controlled by $\left| u \phi_u \right|$, and $\sqrt{R/|u|} \phi_R$. So we just need to control the factor $2mR\phi$. The following Poincar\'e-type estimate and its corollary allow us to do so. They in fact prove more than is strictly necessary for this step, but this will be useful for higher order estimates. We first need to introduce some notation.
\begin{definition} \label{SurfaceBits}
We denote for $0\leq s \leq 1$,
\[ {\cal H}_{s,{u_0}} := {\cal H}_{s} \cap \{ u<u_0 \} \, ,\]
for $s=0$, we use the alternative notation
\[ \scri^+_{u_0} := \scri^+ \cap \{ u<u_0 \} \, . \]
For $0\leq s_0 \leq 1$, in addition to the hypersurface ${\cal H}_{s_0,{u_0}}$, we also consider the part of ${\cal S}_{u_0}$ in the past of ${\cal H}_{s_0}$~:
\[ {\cal S}_{u_0,s_0} = {\cal S}_{u_0} \cap \{ s_0 < s <1 \} \, .\]
\end{definition}
\begin{lemma} \label{SobolEst}
Given $u_0 <0$, there exists a constant $C>0$ such that for any $f\in {\cal C}^\infty_0 (\R )$, we have
\[ \int_{-\infty}^{u_0} \left( f (u) \right)^2 \d u \leq C \int_{-\infty}^{u_0} u^2 \left( f ' (u) \right)^2 \d u \, .\]
\end{lemma}
This has an important consequence\footnote{Not as immediate as one may think. Indeed, $\phi_u$ is the partial derivative of $\phi$ with respect to the variable $u$, it differs from $\frac{\d }{\d u} \left( \phi_{|_{{\cal H}_s}} \right)$ (see equation (\ref{DuPhiHs})).}.
\begin{corollary} \label{CorolL2}
For $u_0 <0$, $|u_0|$ large enough, there exists a constant $C>0$ such that for any smooth compactly supported initial data at $t=0$, the associated rescaled solution $\phi$ satisfies, 
\[ \forall s\in [0,1] \, ,~ \int_{{\cal H}_{s,u_0}} \phi^2 \d u \d^2 \omega \leq  C {\cal E}_{{\cal H}_{s,u_0}} (\phi ) \, . \]
\end{corollary}
From this, we infer the first fundamental estimate.
\begin{theorem} \label{ThmEnEstH1}
For $u_0<0$, $|u_0|$ large enough, there exists a constant $C>0$ such that, for any smooth compactly supported initial data at $t=0$, the associated rescaled solution $\phi$ satisfies for all $0\leq s_0 \leq 1$,
\[ {\cal E}_{{\cal H}_{s_0,{u_0}}} (\phi) \leq C {\cal E}_{{\cal H}_{1,{u_0}}} (\phi ) \, ,\]
and in particular, for $s_0 = 0$,
\[ {\cal E}_{\scri^+_{u_0}} (\phi )\leq C {\cal E}_{{\cal H}_{1,{u_0}}} (\phi ) \, . \]
\end{theorem}
We also prove a converse inequality in order to ensure the optimality of our estimates.
\begin{theorem} \label{ThmConverseH1}
For $u_0 <0$, $|u_0|$ large enough, there exists a constant $C>0$ such that, for any smooth compactly supported initial data at $t=0$, the associated rescaled solution $\phi$ satisfies for all $0\leq s_0 \leq 1$,
\[ {\cal E}_{{\cal H}_{s_0,u_0}} (\phi ) \leq C \left( {\cal E}_{\scri_{u_0 }^+} (\phi ) + {\cal E}_{{\cal S}_{u_0 ,s_0}} (\phi ) \right) \]
and in particular, for $s_0=1$,
\[ {\cal E}_{{\cal H}_{1,u_0}} (\phi ) \leq C \left( {\cal E}_{\scri_{u_0 }^+} (\phi ) + {\cal E}_{{\cal S}_{u_0}} (\phi ) \right) \, . \]
\end{theorem}
Using the foliation $\{ {\cal S}_u \}_u$ and the results of theorem \ref{ThmEnEstH1}, we can prove another fundamental estimate that gives control over the $4$-volume local $L^2$ norm of $\phi_R$ in $\Omega^+_{u_0}$, it is analogous to the point of view on peeling developed in remark \ref{Mink4VolPeeling} for Minkowski space.
\begin{theorem} \label{EnEstDR}
Let $\phi$ be a smooth solution of (\ref{ConfWaveEq}) associated with smooth compactly supported data at $t=0$, we have for all $u\leq u_0$,
\[ \int_{{\cal S}_u} \phi_R^2 \d R \d^2 \omega \leq C {\cal E}_{{\cal H}_{1,u_0}} (\phi )  \, , \]
where the constant $C>0$ is independent of $\phi$ and $u$. In particular, for any compact subset $K$ of $\bar{\Omega}_{u_0}^+$, where $\bar{\Omega}_{u_0}^+ = \Omega_{u_0}^+ \cup \scri^+_{u_0} \cup {\cal H}_{1,u_0} \cup {\cal S}_{u_0}$, there exists a constant $C_K >0$ independent of $\phi$ such that
\[ \int_K \phi_R^2 \d u \d R \d^2 \omega \leq C_K {\cal E}_{{\cal H}_{1,u_0}} (\phi )  \, . \]
\end{theorem}
\begin{remark}
The estimates of theorems \ref{ThmEnEstH1} and \ref{ThmConverseH1} allow us to solve the Goursat problem on $\scri$ for equation (\ref{ConfWaveEq}) and therefore entail a complete conformal scattering theory as defined in \cite{MaN04}.
\end{remark}

\section{Higher order estimates and peeling} \label{Peeling}

In order to obtain estimates on the higher order derivatives of $\phi$, we commute differential operators into equation (\ref{ConfWaveEq}). Because of the symmetries of Schwarzschild's space-time, we have for any $k \in \N$
\begin{gather}
\left( \square + 2mR \right) \partial_u^k \phi = 0 \, , \label{EqDuk} \\
\left( \square + 2mR \right) \nabla_{S^2}^k \phi = 0 \, . \label{EqGradS2k}
\end{gather}
Equations (\ref{EqDuk}) and (\ref{EqGradS2k}) give us on $\partial_u^k \phi$ and $\nabla_{S^2}^k \phi$ respectively, exactly the same energy estimates that we have obtained on $\phi$.

We now need some control over the derivatives of $\phi$ with respect to $R$. Mimicking the proof of the peeling in flat space given in section \ref{MinkPeel} would lead to the use of the Morawetz vector field for this purpose. However, we quickly realize that this leads to serious problems in the Schwarzschild case. Indeed, we have
\begin{eqnarray*}
\left[ T , 2mR \right] &=& - 4m (1+uR) \, ,\\
\left[ T , \left[ T , 2mR \right] \right] &=& 4mu (2+uR)\, ,\\
\left[ T , \left[ T , \left[ T , 2mR \right] \right] \right] &=& 0 \, .
\end{eqnarray*}
Although the third commutator is zero, the second blows up near $i^0$, which leads to error terms that cannot be controlled as soon as we reach two orders of differentiation (other terms with similar behaviour will come from repeated commutations of $T$ with the d'Alembertian).

Instead, we commute the operator $\partial_R$ into the equation to obtain
\begin{equation} \label{EqDR}
\left( \square +2mR \right) \phi_R = 2 (1-3m) R\partial_R \phi_R -2 (1-6mR) \phi_R -2m \phi \, .
\end{equation}
We obtain the following approximate conservation law for $\psi :=\phi_R$~:
\begin{eqnarray}
\nabla^a \left( T^b T_{ab} (\psi) \right) &=& \square \psi T^b\partial_b \psi + \nabla^a T^b T_{ab} (\psi ) \nonumber \\
&=& \left( 2 (1-3m) R\partial_R \psi +2 (1-5mR) \psi -2m \phi \right) T^b\partial_b \psi \nonumber \\
&&+ 4mR^2(3+uR)\psi^2 \label{ApproxConsLawPhiR} \, .
\end{eqnarray}
Using corollary \ref{CorolL2} for $\phi_R$ as well as $\phi$ and following the proof of theorems \ref{ThmEnEstH1} and \ref{ThmConverseH1}, we obtain~:
\begin{proposition} \label{EnEstPhiR}
There exists a constant $C>0$ such that, for any solution $\phi$ of (\ref{ConfWaveEq}) associated to smooth compactly supported initial data, we have for all $0\leq s_0 \leq 1$,
\begin{eqnarray*}
{\cal E}_{{\cal H}_{s_0,u_0}} (\phi_R ) &\leq & C \left( {\cal E}_{{\cal H}_{1,u_0}} (\phi ) + {\cal E}_{{\cal H}_{1,u_0}} (\phi_R ) \right) \, .\\
{\cal E}_{{\cal H}_{s_0,u_0}} (\phi_R ) &\leq & C \left( {\cal E}_{\scri_{u_0 }^+} (\phi ) + {\cal E}_{{\cal S}_{u_0 ,s_0}} (\phi ) + {\cal E}_{\scri_{u_0 }^+} (\phi_R ) + {\cal E}_{{\cal S}_{u_0 ,s_0}} (\phi_R )\right) \, .
\end{eqnarray*}
\end{proposition}
Further commutations of $\partial_R$ into equation (\ref{ConfWaveEq}) will always lead to controllable terms, simply because the coefficients of the error terms are polynomials in $R$. We therefore have the general result~:
\begin{theorem} \label{ThmHigherOrderEst}
For each $k \in \N$, there exists a constant $C_k>0$ such that, for any solution $\phi$ of (\ref{ConfWaveEq}) associated to smooth compactly supported initial data, we have for all $0\leq s_0 \leq 1$,
\begin{eqnarray*}
{\cal E}_{{\cal H}_{s_0,u_0}} (\partial_R^k \phi ) &\leq & C_k \sum_{p=0}^k {\cal E}_{{\cal H}_{1,u_0}} (\partial_R^p \phi ) \, ,\\
{\cal E}_{{\cal H}_{s_0,u_0}} (\partial_R^k \phi ) &\leq & C_k \sum_{p=0}^k \left( 
{\cal E}_{\scri_{u_0}^+} (\partial_R^p \phi ) + {\cal E}_{{\cal S}_{u_0 ,s_0}} (\partial_R^p \phi ) \right)
\end{eqnarray*}
and in particular
\begin{eqnarray*}
{\cal E}_{{\scri^+_{u_0}}} (\partial_R^k \phi ) &\leq & C_k \sum_{p=0}^k {\cal E}_{{\cal H}_{1,u_0}} (\partial_R^p \phi ) \, ,\\
{\cal E}_{{\cal H}_{1,u_0}} (\partial_R^k \phi ) &\leq & C_k \sum_{p=0}^k \left( 
{\cal E}_{\scri_{u_0}^+} (\partial_R^p \phi ) + {\cal E}_{{\cal S}_{u_0}} (\partial_R^p \phi ) \right) \, .
\end{eqnarray*}
\end{theorem}
An similarly to theorem \ref{EnEstDR}, we can prove the following result~:
\begin{theorem} \label{ThmHighEnEstDr}
Let $\phi$ be a smooth solution of (\ref{ConfWaveEq}) associated with smooth compactly supported data at $t=0$. For all $k \in \N$ and $u\leq u_0$, there exists a constant $C_k>0$ independent of $\phi$ and $u$ such that
\[ \int_{{\cal S}_u} \left| \partial_R^k \phi \right|^2 \d R \d^2 \omega \leq C_k \sum_{p=0}^k {\cal E}_{{\cal H}_{1,u_0}} ( \partial_R^p \phi )  \, . \]
In particular, for any compact subset $K$ of $\bar{\Omega}_{u_0}^+$, where $\bar{\Omega}_{u_0}^+ = \Omega_{u_0}^+ \cup \scri^+_{u_0} \cup {\cal H}_{1,u_0} \cup {\cal S}_{u_0}$, there exists a constant $C_{K,k} >0$ such that
\[ \int_K \left| \partial_R^k \phi \right|^2 \d u \d R \d^2 \omega  \leq C_{K,k} \sum_{p=0}^k {\cal E}_{{\cal H}_{1,u_0}} (\partial_R^p \phi )  \, . \]
\end{theorem}
We can now define the function spaces of initial data that guarantee peeling at order $k$ for solutions of (\ref{ConfWaveEq}). First we need to be clear about what we mean by peeling at a certain order.
\begin{definition} \label{PeelingSolution}
We say that a solution $\phi$ of (\ref{ConfWaveEq}) peels at order $k\in \N$ if for all polynomial $P$ in $\partial_R$ and $\nabla_S^2$ of order lower than or equal to $k$, we have ${\cal E}_{\scri^+_{u_0} } (P \phi ) <+\infty$. This means than for all $p \in \{ 0,1,...,k\}$ we have for all $q \in \{ 0,1,...,p \}$, ${\cal E}_{\scri^+_{u_0} } (\partial_R^q \nabla_{S^2}^{p-q} \phi ) <+\infty$.
\end{definition}
By theorem \ref{ThmHigherOrderEst}, the condition on initial data that guarantees peeling at order $k$ is therefore that
\[ \forall p \in \{ 0,1,...,k\} \, ,~ \forall q \in \{ 0,1,...,p \} \, , ~ {\cal E}_{{\cal H}_{1,u_0} } (\partial_R^q \nabla_{S^2}^{p-q} \phi ) <+\infty \, .\]
This can be expressed as a condition purely on the initial data that no longer involves the full solution $\phi$. First note that equation (\ref{ConfWaveEq}) in terms of variables $t,r_*,\omega$ becomes
\[ \left( \partial_t^2 - \partial_{r_*}^2 + F \left( \frac{2m}{r^3} - \frac{1}{r^2} \Delta_{S^2} \right) \right) \phi = 0 \, ,~\mbox{where } F(r) := 1-\frac{2m}{r} \, .\]
Whence,
\begin{eqnarray*}
\partial_R \left( \begin{array}{c} \phi \\ \partial_t \phi \end{array} \right) &=& -\frac{r^3}{r-2m} \left( \partial_t + \partial_{r_*} \right) \left( \begin{array}{c} \phi \\ \partial_t \phi \end{array} \right) \\
&=& -\frac{r^3}{r-2m} \left( \begin{array}{cc} {\partial_{r_*}} & 1 \\ {\partial_{r_*}^2 - \frac{2mF}{r^3} + \frac{F}{r^2} \Delta_{S^2}} & {\partial_{r_*}} \end{array} \right) \left( \begin{array}{c} \phi \\ \partial_t \phi \end{array} \right) =: L \left( \begin{array}{c} \phi \\ \partial_t \phi \end{array} \right) \, .
\end{eqnarray*}
The operator $L$ purely involves spacelike derivatives. We can now express the spaces of initial data that entail peeling at a given order.
\begin{definition} \label{PeelingSpaces}
Given $\phi_0,\phi_1 \in {\cal C}^\infty_0 (\left[ -u_0 ,+\infty \right[_{r_*} \times S^2_\omega )$, we define the following squared norm of order $k$~:
\begin{equation}
\left\| \left( \begin{array}{c} \phi \\ \partial_t \phi \end{array} \right) \right\|_k^2 := \sum_{p=0}^k \sum_{q=0}^p {\cal E}_{{\cal H}_{1,u_0}} \left( L^q \nabla_{S^2}^{p-q} \left( \begin{array}{c} \phi_0 \\ \phi_1 \end{array} \right) \right) \, , \label{OrderkNorm}
\end{equation}
where we have denoted by ${\cal E}_{{\cal H}_{1,u_0}} \left( \begin{array}{c} \phi_0 \\ \phi_1 \end{array} \right)$ the energy ${\cal E}_{{\cal H}_{1,u_0}} (\phi )$ where $\phi$ is replaced by $\phi_0$ and $\partial_t \phi = \phi_u$ is replaced by $\phi_1$.

The space of initial data (on $\left[ -u_0 ,+\infty \right[_{r_*} \times S^2_\omega$) for which the associated solution peels at order $k$ is the completion of ${\cal C}^\infty_0 (\left[ -u_0 ,+\infty \right[_{r_*} \times S^2_\omega ) \times {\cal C}^\infty_0 (\left[ -u_0 ,+\infty \right[_{r_*} \times S^2_\omega )$ in the norm (\ref{OrderkNorm}). The fact that we have estimates both ways at all orders in theorem (\ref{ThmHigherOrderEst}) guarantees that this setting is optimal in our framework.
\end{definition}
\begin{remark}
It is important to note that all estimates in this paper are locally uniform in $m$. In particular, $u_0$ and all the constants in all the theorems can be fixed independently of $m$ for $m$ in a fixed compact domain $[0,M]$.
\end{remark}
\begin{remark}
Similarly, for two different values of the mass $m$, the function spaces of initial data for a given regularity are canonically isomorphic. The equivalence in the norms is uniform for $m$ in a fixed compact domain $[0,M]$. This is because the quantity $r_*-r$ never appears in these norms. Hence, we do not see any logarithmic divergence between the norms for two different values of $m$.
\end{remark}

\section{Interpretation of the results} \label{Interpretation}

The characterization of the peeling given in definition \ref{PeelingSpaces} proves the existence of large classes of data that guarantee peeling at any given order on the Schwarzschild space-time. It is interesting to see whether these classes are natural extensions to the Schwarzschild case of the classes of data for Minkowski space-time given in definition \ref{DefPeelMinkEnEst}.

For $m=0$, all our estimates proved for Schwarzschild's space-time are still valid and give the corresponding estimates for Minkowski space-time. Accordingly, definition \ref{PeelingSpaces} for $m=0$ gives a fourth definition of the peeling in flat space-time that is as follows~:
\begin{definition} \label{DefPeelMinkSchwaNoMass}
Given $\phi_0,\phi_1 \in {\cal C}^\infty_0 (\left[ -u_0 ,+\infty \right[_r \times S^2_\omega )$, we define the following squared norm of order $k$~:
\begin{equation}
\left\| \left( \begin{array}{c} \phi \\ \partial_t \phi \end{array} \right) \right\|_k^2 := \sum_{p=0}^k \sum_{q=0}^p {\cal E}_{{\cal H}_{1,u_0}} \left( L^q \nabla_{S^2}^{p-q} \left( \begin{array}{c} \phi_0 \\ \phi_1 \end{array} \right) \right) \, , \label{OrderkNormMink}
\end{equation}
where
\[ L = -r^2 \left( \begin{array}{cc} {\partial_{r}} & 1 \\ {\partial_{r}^2 + \frac{1}{r^2} \Delta_{S^2}} & {\partial_{r}} \end{array} \right) \left( \begin{array}{c} \phi \\ \partial_t \phi \end{array} \right) \, .\]

The space of initial data (on $\left[ -u_0 ,+\infty \right[_r \times S^2_\omega$) for which the associated solution peels at order $k$ is the completion of ${\cal C}^\infty_0 (\left[ -u_0 ,+\infty \right[_r \times S^2_\omega ) \times {\cal C}^\infty_0 (\left[ -u_0 ,+\infty \right[_r \times S^2_\omega )$ in the norm (\ref{OrderkNormMink}).
\end{definition}
Of course definitions \ref{DefPeelMinkEnEst} and \ref{DefPeelMinkSchwaNoMass} are of a different nature since the former considers spaces of data on the whole $\{ t=0 \}$ slice whereas the latter gives only data in a neighbourhood of $i^0$. However, we can compare the spaces of data on the domain $[ -u_0 ,+\infty [_r \times S^2_\omega $. At the level of basic energies, we show that the two definitions are equivalent. First, we introduce some notations so that we can clearly identify the physical solution and the rescaled fields obtained using $\Omega$ and $R$. We shall denote the physical field, i.e. the solution of (\ref{MinkWaveEq}) on $\mathbb{M}$, by $\tilde{\phi}$. The field rescaled using the conformal factor $\Omega$ defined in (\ref{FullConfFact}) will be denoted by $\psi := \Omega^{-1} \tilde{\phi}$ and when rescaling using $R =1/r$, we shall use the notation $\phi := R^{-1} \tilde{\phi} = r \tilde{\phi}$.
\begin{proposition} \label{DiffEnergyFrameworks}
In terms of $\tilde{\phi}$, the energy at $\tau =0$ on the Einstein cylinder reads~:
\begin{gather}
{\cal E}_{X_0} (\psi ) = \int_{\{ t=0 \}} \left( \tilde{\phi}_t^2 + \left( \tilde{\phi}_r + \frac{2r}{1+r^2} \tilde{\phi} \right)^2 + \left| \frac{\nabla_{S^2} \tilde{\phi}}{r} \right|^2 + \frac{4\tilde{\phi}^2}{(1+r^2)^2} \right) \frac{1+r^2}{4} r^2 \d r \d^2 \omega \label{ControlEDTau} \\
\simeq \int_{\Sigma_0} \left( (1+r^2) \left( \partial_t \tilde{\phi} \right)^2 + (1+r^2) \left( \partial_r \tilde{\phi} \right)^2 + (1+r^2) \left| \frac{\nabla_{S^2} \tilde{\phi}}{r} \right|^2 + \tilde{\phi}^2 \right) r^2 \d r \d^2 \omega \, ; \label{EqControl1Control2}
\end{gather}
while the energy at $t=0$ for $\phi$ satisfies~:
\begin{equation} \label{ControlEMorawetz}
{\cal E}_{{\cal H}_{1,u_0}} (\phi ) \simeq \int_{\{ t=0 \, ,~ r > -u_0 \}} \left\{ r^2 \tilde{\phi}_t^2 + r^2 \tilde{\phi}_r^2 + r^2 \left| \frac{\nabla_{S^2} \tilde{\phi }}{r} \right|^2 + \tilde{\phi}^2 \right\} r^2 \d r \d^2 \omega \, .
\end{equation}
\end{proposition}

It is for higher orders that the definitions differ. On the Einstein cylinder, we raise regularity with $\partial_\tau$, which, on the hypersurface $\{ t=0 \}$, coincides with $\frac{1+ r^2}{2} \partial_t$. As we can see from identities (\ref{EinsteinEnEqEven}) and (\ref{EinsteinEnEqOdd}), the application of $\partial_\tau$ amounts to that of the elliptic operator $\left( 1-\Delta_{S^3} \right)^{1/2}$, which can be understood, thanks to the ellipticity, as providing a control over $r^2 \partial_r$ and angular derivatives on $S^2$ independently. In the partial compactification obtained when rescaling with $R$, the vector we use for raising regularity in the estimates is $\partial_R = -r^2 (\partial_t + \partial_r )$~; it is a null vector field whose action cannot be understood as that of an elliptic operator. Although it also acts as a combination of $r^2 \partial_r$ and angular derivatives, it gives a weaker control on initial data.

So, our characterization of the peeling on Schwarzchild's space-time is a natural generalization of a definition on Minkowski space that is not equivalent to the usual description of the peeling~: it provides larger classes of data giving rise to solutions that peel at a certain order. This is because we pinpoint the relevant null derivative to control near $i^0$ instead of controlling all derivatives. Our paper thus not only proves the validity of the peeling model for the wave equation on the Schwarzschild metric at all orders, it also provides a description of the peeling on flat space-time that is more general than the one used since Penrose's paper in 1963.

\appendix

\section{Proofs of the main results} \label{Proofs}

The following lemma contains trivial but essential observations~:
\begin{lemma} \label{ApproxCloseI0}
Let $\varepsilon >0$, then for $u_0<0$, $|u_0|$ large enough, in the domain $\Omega_{u_0}^+$, we have
\[ r< r_* <r(1+\varepsilon ) \, ,~ 1 < Rr_* < 1+\varepsilon  \, ,~ 0 < R|u| < 1+\varepsilon  \, ,~ 1-\varepsilon < 1-2mR <1 \, ,\]
and of course
\[ 0< s=\frac{|u|}{r_*} < 1 \, . \]
\end{lemma}

\subsection{Proof of lemma \ref{LemmaNormeEquivEnergy}}

The restriction of the $3$-form $T^aT_{ab}\d^3x^b$ to the hypersurface ${\cal H}_s$ is given by
\begin{gather*}
\left. T^aT_{ab}\d^3x^b \right|_{{\cal H}_s} \\
= \left\{ u^2 \phi_u^2 + R^2 (1-2mR) \left( \frac{(2+uR)^2}{2s} - 1 -uR - \frac{mu^2R^3}{s} \right) \phi_R^2 \right. \\
 \left. + R^2 (1-2mR) u^2 \phi_u \phi_R + \left( \frac{R^2 u^2(1-2MR)}{2s}
+1+uR \right) \left| \nabla_{S^2} \phi \right|^2 \right\} \d u
\wedge \d^2\omega
\end{gather*}
and using the definition of ${\cal H}_s = \{ u=-sr_* \}$, we get
\begin{gather*}
\left. T^aT_{ab}\d^3x^b \right|_{{\cal H}_s} \\
= \left\{ u^2 \phi_u^2 + R^2 (1-2mR) \left( \frac{(2+uR)^2}{2} \frac{r_*}{|u|} - 1 -uR - m|u|R^3r_* \right) \phi_R^2 \right. \\
\left. + R^2 (1-2mR) u^2 \phi_u \phi_R + \left( \frac{R^2 (1-2MR)}{2} |u|r_* +1+uR \right) \left| \nabla_{S^2} \phi \right|^2 \right\} \d u \wedge \d^2\omega
\end{gather*}
Let $\varepsilon >0$ small, we choose $u_0 <<-1$ such as in lemma \ref{ApproxCloseI0}.
We first deal with the angular term~:
\begin{eqnarray*}
\frac{R^2 (1-2MR)}{2} |u|r_* +1+uR &=& 1 + |u|R \left( -1 + \frac{1}{2} Rr_* (1-2mR) \right) \\
&=&  1 + |u|R \left( -1 + \frac{1}{2} (1+2m R\log (r-2m)) (1-2mR) \right) \\
&=& 1 + |u|R \left( -\frac{1}{2} + mR \left( -1 + (1-2mR) \log (r-2m) \right) \right)
\end{eqnarray*}
and taking $|u_0|$ large enough, we get
\[ 1\geq \frac{R^2 (1-2MR)}{2} |u|r_* +1+uR \geq \frac{1}{2} (1-\varepsilon ) \, .\]
Now we estimate above and below the term in $\phi_R^2$~:
\begin{gather*}
R^2 (1-2mR) \left( \frac{(2+uR)^2}{2} \frac{r_*}{|u|} - 1 -uR - m|u|R^3r_* \right) \\
= \frac{R}{|u|} (1-2mR) \left( \frac{(2+uR)^2}{2} Rr_* - |u|R +(uR)^2 - mR(uR)^2Rr_* \right) \\
= \frac{R}{|u|} (1-2mR) \left( \left( 2+2uR + \frac{(uR)^2}{2} \right) Rr_* +uR + (uR)^2 - mR(uR)^2Rr_* \right) \\
= \frac{R}{|u|} (1-2mR) \left( 2+2uR + \frac{(uR)^2}{2} +uR + (uR)^2 \right) \\
+ \frac{R}{|u|} (1-2mR) \left( \left( 2+2uR + \frac{(uR)^2}{2} \right) 2mR \log (r-2m) - mR(uR)^2Rr_* \right)
\end{gather*}
By lemma \ref{ApproxCloseI0}, the term in the penultimate line above dominates the term in the last line. It can be written as
\[ \frac12 \frac{R}{|u|} (1-2mR) \left( 3 (uR)^2 + 6 uR +4 \right) \, . \]
Recall that $-1-\varepsilon < Ru <0$ in $\Omega_{u_0}^+$. Noting that the function $f(x) = 3x^2+6x+4$ is everywhere positive on $\R$, takes its minimum at $x=-1$, $f(-1) = 1$ and $f(x) \in [1,4]$ for $x\in [-2,0]$, we infer that for $|u_0|$ large enough, $u_0<0$, we have in $\Omega_{u_0}^+$ say
\[ \frac{1}{3} \frac{R}{|u|} \leq R^2 (1-2mR) \left( \frac{(2+uR)^2}{2} \frac{r_*}{|u|} - 1 -uR - m|u|R^3r_* \right) \leq 5 \frac{R}{|u|} \, .\]
Then we treat the term in $\phi_u \phi_R$ in the following way~:
\begin{eqnarray*}
\left| R^2 (1-2mR) u^2 \phi_u \phi_R \right| &\leq & (R|u|)^{3/2} \left| u \phi_u \right| \left| \sqrt{\frac{R}{|u|}}\, \phi_R \right| \\
&\leq & (1+\varepsilon )^{3/2} \left| u \phi_u \right| \left| \sqrt{\frac{R}{|u|}}\, \phi_R \right| \\
&\leq & (1+\varepsilon )^{3/2} \frac{1}{2} \left( \lambda^2 u^2 \phi_u^2 + \frac{1}{\lambda^2} \frac{R}{|u|} \phi_R^2 \right) \, , ~\lambda \in \R^* \, .
\end{eqnarray*}
This guarantees that the energy density on ${\cal H}_s$ is estimated above by (\ref{NormeEquivEnergy}) uniformly in $\Omega_{u_0}^+$. Now in order to establish the estimate below, we need to choose $\lambda$ such that
\[ \frac{\lambda^2}{2} < 1~\mathrm{and~} \frac{1}{2\lambda^2} < \frac{1}{3} \, , \]
i.e. $\sqrt{3/2}<\lambda <\sqrt{2}$. Taking for example $\lambda = \frac{1}{2} (\sqrt{3/2}+\sqrt{2})$, this concludes the proof of the lemma, provided $\varepsilon $ is small enough. \qed

\subsection{Proof of lemma \ref{SobolEst}}

This is a simple integration by parts~:
\[ \int_{-\infty}^{u_0} f^2 \d u = \left[ (u-u_0 ) f^2 \right]_{-\infty}^{u_0} - \int_{-\infty}^{u_0} (u-u_0 ) 2 ff ' \d u \]
The boundary term vanishes (recall that $f$ is assumed to be compactly supported, which gets rid of the boundary term at $-\infty$) and using $u_0 <0$, we get
\begin{eqnarray*}
\int_{-\infty}^{u_0} f^2 \d u &\leq & \int_{-\infty}^{u_0} |u-u_0| 2 |f | |f '| \d u \leq \int_{-\infty}^{u_0} 2 |f | |uf '| \d u\\
& \leq &\frac{1}{2} \int_{-\infty}^{u_0} f^2 \d u + 2 \int_{-\infty}^{u_0} u^2 \left( f ' \right)^2 \d u \, ,
\end{eqnarray*}
which gives the result. \qed

\subsection{Proof of corollary \ref{CorolL2}}

We simply need to show that
\[ \int_{{\cal H}_{s,u_0}} \left| u \frac{\d }{\d u} \left( \phi_{|_{{\cal H}_s}} \right) \right|^2 \d u \d^2 \omega \]
is controlled uniformly by ${\cal E}_{{\cal H}_{s,u_0}} (\phi )$. For a given $s\in [0,1]$, we have
\[ \phi_{|_{{\cal H}_s}} (u, \omega ) = \phi \left( u , R= \frac{1}{G^{-1} ( -u/s )} , \omega \right) \, ,\]
where $G$ is the function  $r\mapsto r_*=G(r)$ (see (\ref{FunctionRstar})). Hence,
\begin{eqnarray}
\frac{\d }{\d u} \left( \phi_{|_{{\cal H}_s}}  \right) &=& \left. \left( \phi_u -\frac{1}{s} \frac{-1}{G^{-1} ( -u/s )} (G^{-1})' (-u/s)  \phi_R \right) \right|_{{\cal H}_s} \nonumber \\
&=& \left. \left(  \phi_u -\frac{R}{u} \frac{r_*R}{1-2mR} \phi_R \right) \right|_{{\cal H}_s} \, . \label{DuPhiHs}
\end{eqnarray}
By lemma \ref{ApproxCloseI0}, it follows that
\[ \int_{{\cal H}_{s,u_0}} \left| u \frac{\d }{\d u} \left( \phi_{|_{{\cal H}_s}} \right) \right|^2 \d u \d^2 \omega \lesssim \int_{{\cal H}_{s,u_0}} \left( u^2 \phi_u^2 + \frac{R}{|u|} \phi_R^2 \right) \d u \d^2 \omega \lesssim {\cal E}_{{\cal H}_{s,u_0}} (\phi )\, . \]
This concludes the proof. \qed

\subsection{Proof of theorem \ref{ThmEnEstH1}}

The surfaces ${\cal H}_{s,{u_0}}$, $s_0 \leq s \leq 1$, parametrized
by $\tau$, will be denoted $\Sigma_{\tau , u_0}$, $0 \leq \tau \leq
\tau (s_0 )$. We also consider ${\cal S}_{u_0}^{s_0}$ which is the
part of ${\cal S}_{u_0}$ situated in the past of ${\cal H}_{s_0}$,
i.e. for which\footnote{This hypersurface is not the part of ${\cal
    S}_{u_0}$ introduced in definition \ref{SurfaceBits} and
  considered in theorem \ref{ThmConverseH1}; instead, up to a negligable set, it is the complement of ${\cal S}_{u_0,s_0}$ in ${\cal S}_{u_0}$.} we have $s_0<s<1$. We denote by $\mathit{Err}$ the total error term on each hypersurface ${\cal H}_s$
\begin{eqnarray*}
\mathit{Err} &:=& 4mR^2(3+uR) \phi_R^2 (r_* R)^{3/2} (1-2mR) \sqrt{\frac{R}{|u|}} \\
&& -2mR \phi \left( u^2 \partial_u \phi - 2 (1+uR) \partial_R \phi \right) (r_* R)^{3/2} (1-2mR) \sqrt{\frac{R}{|u|}} \, .
\end{eqnarray*}
This is estimated uniformly on $\Omega_{u_0}^+$ by
\[ \left| \mathit{Err} \right| \lesssim \frac{R}{|u|} \phi_R^2 + |\phi | |u\phi_u | + |\phi | \sqrt{\frac{R}{|u|}} |\phi_R | \lesssim u^2 \phi_u^2 + \frac{R}{|u|} \phi_R^2 + \phi^2 \, . \]
We use the approximate conservation law (\ref{ApproxConsLaw}) on the domain $\Omega_{u_0}^+ \cap \left\{ s_0<s< 1 \right\}$ to obtain
\begin{eqnarray*}
&& {\cal E}_{{\cal H}_{s_0 , u_0}} (\phi )+ {\cal E}_{{\cal S}_{u_0}^{s_0}} (\phi ) - {\cal E}_{{\cal H}_{1,{u_0}}} (\phi ) = \int_0^{\tau (s_0)} \int_{{\Sigma}_{\tau,u_0}} \mathit{Err} \, \d \tau \, \d u \, \d^2 \omega \\
&& \hspace{1in} \lesssim \int_0^{\tau (s_0)} {\cal E}_{{\Sigma}_{\tau,{u_0}}} (\phi ) \d \tau + \int_0^{\tau (s_0)} \left( \int_{{\Sigma}_{\tau,u_0}} \phi^2 \d u \d^2 \omega \right) \d \tau \, .
\end{eqnarray*}
Furthermore, since $(2+uR)^2 - 2mu^2R^3 \geq 0$ in $\Omega^+_{u_0}$, we have by (\ref{EnergySu}) that ${\cal E}_{{\cal S}_{u_0}^{s_0}} (\phi ) \geq 0$. Thus, using lemma \ref{SobolEst}, we get
\begin{eqnarray*}
{\cal E}_{{\Sigma}_{\tau (s_0) ,u_0}} (\phi ) &\leq & {\cal E}_{{\Sigma}_{\tau (s_0) ,u_0}} (\phi ) + {\cal E}_{{\cal S}_{u_0, s_0}} (\phi ) \\
&\lesssim & {\cal E}_{{\Sigma}_{0,{u_0}}} (\phi ) + \int_{0}^{\tau (s_0 )} {\cal E}_{{\Sigma}_{\tau (s),{u_0}}} (\phi ) \d \tau + \int_0^{\tau (s_0)} \left( \int_{{\Sigma}_{\tau,u_0}} \phi^2 \d u \d^2 \omega \right) \d \tau \\
&\lesssim & {\cal E}_{{\Sigma}_{0,{u_0}}} (\phi ) + \int_{0}^{\tau (s_0 )} {\cal E}_{{\Sigma}_{\tau (s),{u_0}}} (\phi ) \d \tau  \, .
\end{eqnarray*}
The result follows via a standard Gronwall estimate. \qed

\subsection{Proofs of proposition \ref{EnEstPhiR} and theorems \ref{ThmHigherOrderEst} and \ref{ThmHighEnEstDr}}

They are straightforward extensions by induction of the proofs of theorems \ref{ThmEnEstH1}, \ref{ThmConverseH1} and \ref{EnEstDR}.

\subsection{Proof of theorem \ref{ThmConverseH1}}

It is very similar to that of theorem \ref{ThmEnEstH1}. On the domain $\Omega_{u_0}^+ \cap \left\{ 0<s<s_0 \right\}$, we use the approximate conservation law (\ref{ApproxConsLaw})
\[ {\cal E}_{\scri^+_{u_0}} (\phi ) + {\cal E}_{{\cal S}_{u_0, s_0}}
(\phi ) - {\cal E}_{{\Sigma}_{\tau (s_0 ),{u_0}}} (\phi ) = \int_{\tau
  (s_0 )}^2 \int_{{\Sigma}_{\tau (s) , u_0}} \mathit{Err} \,\, \d u \d^2 \omega \d \tau \, . \]
We obtain
\[ {\cal E}_{{\Sigma}_{\tau (s_0 ),{u_0}}} (\phi ) \lesssim  {\cal E}_{\scri^+_{u_0}} (\phi ) + {\cal E}_{{\cal S}_{u_0, s_0}} (\phi ) + \int_{\tau (s_0 )}^2 {\cal E}_{{\Sigma}_{\tau (s),{u_0}}} (\phi ) \d \tau \, . \]
The result follows via a standard Gronwall estimate. \qed

\subsection{Proof of theorem \ref{EnEstDR}}

Given $u_1 <u_0$, we use the approximate conservation law (\ref{ApproxConsLaw}) on the domain
\[ \Omega_{u_1}^+ = \left\{ 0\leq s \leq 1 \right\} \cap \left\{ u < u_1 \right\} \, .\]
We obtain
\begin{gather*}
{\cal E}_{\scri^+_{u_1}} (\phi ) + {\cal E}_{{\cal S}_{u_1}} (\phi) - {\cal E}_{{\cal H}_{1,u_1}} (\phi) \\
= \int_{\Omega_{u_1}^+} \left( 4mR^2(3+uR) \phi_R^2 - 2mR \phi \left( u^2 \phi_u - 2(1+uR) \partial_R \phi \right) \right) \d u \, \d R \, \d^2 \omega  \, .
\end{gather*}
Foliating $\Omega_{u_1}^+$ by $\{ {\Sigma}_\tau \}_{\tau ,u_0}$, we
gain in the integral a factor of $(r_* R)^{3/2} (1-2mR) \sqrt{R/|u|}$
and so all the terms are controlled by the energy density on ${\Sigma}_{\tau ,u_1}$ (see proof of theorem \ref{ThmEnEstH1} for details), which is itself controlled uniformly by the energy density of $\Sigma_{0,u_1} = {\cal H}_{1,u_1}$ (applying the result of theorem \ref{ThmEnEstH1} with $u_1$ instead of $u_0$). Whence,
\[ {\cal E}_{\scri^+_{u_1}} (\phi ) + {\cal E}_{{\cal S}_{u_1}} (\phi) - {\cal E}_{{\cal H}_{1,u_1}} (\phi) \lesssim {\cal E}_{{\cal H}_{1,u_1}} (\phi ) \, . \]
Then, the positivity of ${\cal E}_{\scri^+_{u_1}} (\phi)$ gives us the inequality
\[ {\cal E}_{{\cal S}_{u_1}} (\phi) = \int_{{\cal S}_{u_1}} \frac{1}{2} \left[ \left((2+uR)^2 - 2mu^2R^3\right)\phi_R^2+u^2 |\nabla_{S^2}\phi|^2 \right]\d R \, \d ^2 \omega \lesssim {\cal E}_{{\cal H}_{1,u_0}} (\phi) \, .\]
Finally lemma \ref{ApproxCloseI0} entails
\[ \int_{{\cal S}_{u_1}} \phi_R^2 \d R \, \d ^2 \omega \lesssim {\cal E}_{{\cal S}_{u_1}} (\phi) \lesssim {\cal E}_{{\cal H}_{1,u_0}} (\phi) \, ,\]
which concludes the proof. \qed

\subsection{Proof of proposition \ref{DiffEnergyFrameworks}}

First we express $\psi_\tau$ and $\psi_\zeta$ in terms of $\tilde{\phi}_t$ and $\tilde{\phi}_r$. We have
\begin{eqnarray*}
\frac{\partial}{\partial \tau} &=& \frac{1}{2} \left( (1+t^2+r^2) \frac{\partial}{\partial t} + 2tr \frac{\partial}{\partial r} \right) \, , \\
\frac{\partial}{\partial_ \zeta} &=& \frac{1}{2} \left( 2tr \frac{\partial}{\partial t} + (1+t^2+r^2) \frac{\partial}{\partial r}  \right) \, .
\end{eqnarray*}
Hence
\begin{eqnarray*}
\partial_\tau \psi &=& \frac{1+t^2+r^2}{2} \, \Omega^{-1} \left( \frac{t+r}{1+(t+r)^2} + \frac{t-r}{1+(t-r)^2} + \partial_t \right) \tilde{\phi} \\
&& + tr \Omega^{-1} \left( \frac{t+r}{1+(t+r)^2} - \frac{t-r}{1+(t-r)^2} + \partial_r \right) \tilde{\phi} \, , \\
\partial_\zeta \psi &=& tr \Omega^{-1} \left( \frac{t+r}{1+(t+r)^2} + \frac{t-r}{1+(t-r)^2} + \partial_t \right) \tilde{\phi} \\
&& + \frac{1+t^2+r^2}{2} \, \Omega^{-1} \left( \frac{t+r}{1+(t+r)^2} - \frac{t-r}{1+(t-r)^2} + \partial_r \right) \tilde{\phi} \, .
\end{eqnarray*}
In particular at $t=0$ (i.e. at $\tau =0$ as well) we obtain
\begin{eqnarray*}
\partial_\tau \psi (0) &=& \frac{(1+r^2)^2}{4} \partial_t \tilde{\phi} \, , \\
\partial_\zeta \psi (0) &=& \frac{(1+r^2)^2}{4} \left( \frac{2r}{1+r^2}+ \partial_r \right) \tilde{\phi} \, .
\end{eqnarray*}
Besides the Euclidian measure on the $3$-sphere $\{ \tau = 0 \}$ becomes, in terms of variables $(r,\theta , \varphi)$,
\[ \d \mu_{S^3} = \sin^2 \zeta \, \d \zeta \d^2 \omega = \frac{4r^2}{(1+r^2)^2} \frac{2}{1+r^2} \, \d r \d^2\omega \, . \]
\begin{eqnarray*}
{\cal E}_{X_0} (\psi ) &=& \frac{1}{2} \int_{\Sigma_0} \left( \psi_\tau^2 + \left| \nabla_{S^3} \psi \right|^2 + \psi^2 \right) \d \mu_{S^3} \\
&=& \frac{1}{2} \int_{\Sigma_0} \left( \psi_\tau^2 + \psi_\zeta^2 + \frac{1}{\sin^2 \zeta} \left| \nabla_{S^2} \psi \right|^2 + \psi^2 \right) \d \mu_{S^3} \\
&=& \frac{1}{2} \int_{\{ t=0 \}} \left( \frac{(1+r^2)^4}{16} \left( \partial_t \tilde{\phi} \right)^2 + \frac{(1+r^2)^4}{16} \left( \partial_r \tilde{\phi} + \frac{2r}{1+r^2} \tilde{\phi} \right)^2 \right. \\
&& \left. + \frac{(1+r^2)^2}{4r^2} \frac{(1+r^2)^2}{4} \left| \nabla_{S^2} \tilde{\phi} \right|^2 + \frac{(1+r^2)^2}{4} \tilde{\phi}^2 \right) \frac{4r^2}{(1+r^2)^2} \frac{2}{1+r^2} \, \d r \d^2\omega \\
&=& \frac{1}{2} \int_{\{ t=0 \}} \left( \frac{1}{2} (1+r^2) \left( \partial_t \tilde{\phi} \right)^2 + \frac{1}{2} (1+r^2) \left( \partial_r \tilde{\phi} + \frac{2r}{1+r^2} \tilde{\phi} \right)^2 \right. \\
&& \left. + \frac{1}{2} (1+r^2) \frac{1}{r^2} \left| \nabla_{S^2} \tilde{\phi} \right|^2 + \frac{2}{1+r^2} \tilde{\phi}^2 \right) r^2 \d r \d^2 \omega \, .
\end{eqnarray*}
This gives (\ref{ControlEDTau}).

Let us now consider only the two following terms~:
\begin{gather*}
I:= \int_{\Sigma_0} \left( \frac{1}{2} (1+r^2) \left( \partial_r \tilde{\phi} + \frac{2r}{1+r^2} \tilde{\phi} \right)^2 + \frac{2}{1+r^2} \tilde{\phi}^2 \right) r^2 \d r \d^2 \omega \\
= \int_{\Sigma_0} \left( \frac{1+r^2}{2} \tilde{\phi}_r^2 + 2r \tilde{\phi} \tilde{\phi}_r  + \frac{2r^2}{1+r^2} \tilde{\phi}^2 + \frac{2}{1+r^2} \tilde{\phi^2} \right) r^2 \d r \d^2 \omega \\
= \int_{\Sigma_0} \left( \frac{1+r^2}{2} \tilde{\phi}_r^2 + 2r \tilde{\phi} \tilde{\phi}_r  + 2 \tilde{\phi^2} \right) r^2 \d r \d^2 \omega \, .
\end{gather*}
By integration by parts, we have\footnote{Note that $\tilde{\phi}_r$ certainly does not have a limit at $r=0$ in the generic case, but it remains bounded in the neighbourhood of $r=0$, which is enough to justify the integration by parts and prove the equality.}
\[ \int_0^{+\infty} 2r^3 \tilde{\phi} \tilde{\phi}_r \d r = -3 \int_0^{+\infty} \tilde{\phi}^2 r^2 \d r \, ,\]
whence
\[ I= \int_{\Sigma_0} \left( \frac{1+r^2}{2} \tilde{\phi}_r^2 - \tilde{\phi^2} \right) r^2 \d r \d^2 \omega \, . \]
It remains to show that the first term in the above integral compensates for the second and gives us a control over the integral of $\tilde{\phi}^2$. We proceed similarly as for the proof of lemma \ref{SobolEst}~:
\begin{eqnarray*}
\int_0^{+\infty} \tilde{\phi}^2 r^2 \d r &=& -\frac{2}{3} \int_0^{+\infty} r^3 \tilde{\phi} \tilde{\phi}_r \d r \\
&\leq & \frac{\lambda}{3} \int_0^{+\infty} \tilde{\phi}^2 r^2 \d r + \frac{1}{3\lambda} \int_0^{+\infty} \tilde{\phi}_r^2 r^4 \d r \, , ~\mbox{for any } \lambda >0 \, ,
\end{eqnarray*}
and therefore
\[ \lambda (3-\lambda) \int_0^{+\infty} \tilde{\phi}^2 r^2 \d r \leq \int_0^{+\infty} \tilde{\phi}_r^2 r^4 \d r \, . \]
The coefficient $\lambda (3-\lambda )$ is maximum for $\lambda = 3/2$ where it takes the value $9/4$. Hence, for any $\varepsilon >0$
\[ I \geq \int_{\Sigma_0} \left( \varepsilon (1+r^2 ) \tilde{\phi}_r^2 + \left( \frac{1}{2} - \varepsilon \right) \frac{9}{4} \tilde{\phi}^2 - \tilde{\phi}^2 \right) r^2 \d r \d^2 \omega \]
and taking $0< \varepsilon < 1/18$, say $\varepsilon = 1/36$, we obtain
\[ I \geq \int_{\Sigma_0} \left( \frac{1+r^2}{36} \tilde{\phi}_r^2 + \frac{1}{16} \tilde{\phi}^2 \right) r^2 \d r \d^2 \omega \]
from whence we conclude
\[ \int_{\Sigma_0} \left( (1+r^2) \left( \partial_t \tilde{\phi} \right)^2 + (1+r^2) \left( \partial_r \tilde{\phi} \right)^2 + (1+r^2) \frac{1}{r^2} \left| \nabla_{S^2} \tilde{\phi} \right|^2 + \tilde{\phi}^2 \right) r^2 \d r \d^2 \omega \lesssim {\cal E}_{X_0} (\psi ) \, .\]
The reciprocal inequality is trivial. This proves equivalence
(\ref{EqControl1Control2}). 

To prove equivalence (\ref{ControlEMorawetz}), we first use lemma (\ref{LemmaNormeEquivEnergy}) and then the equalities $\phi =r \tilde{\phi}$, $\partial_u = \partial_t$ and $\partial_R = -r^2 \left( \partial_t + \partial_r \right)$. We obtain
\begin{eqnarray*}
{\cal E}_{{\cal H}_{1,u_0}} &\simeq & \int_{\{ r>-u_0 \} } \left( r^2 \tilde{\phi}_t^2 + r^2 \left( \tilde{\phi}_t + \tilde{\phi}_r + \frac{\tilde{\phi}}{r} \right)^2 + r^2 \left| \frac{\nabla_{S^2} \tilde{\phi}}{r} \right|^2 + \tilde{\phi}^2 \right) r^2 \d r \d^2 \omega \\
&& = \int_{\{ r>-u_0 \} } \left( 2r^2 \tilde{\phi}_t^2 + r^2 \tilde{\phi}_r^2 + r^2 \left| \frac{\nabla_{S^2} \tilde{\phi}}{r} \right|^2 + 2\tilde{\phi}^2 \right) r^2 \d r \d^2 \omega \\
&&+ \int_{\{ r>-u_0 \} } \left( 2r^2 \tilde{\phi}_t \tilde{\phi}_r+ 2r \tilde{\phi}_t \tilde{\phi} + 2r \tilde{\phi}_r \tilde{\phi} \right) r^2 \d r \d^2 \omega \, .
\end{eqnarray*}
Now for any $A,B,C \in \R$, we have
\[ 2AB + 2AC + 2BC \leq 2 (A^2+B^2+C^2) \]
and using the facts that
\[ \left( \sqrt{\frac{3}{2}} A + \sqrt{\frac{2}{3}} B + \sqrt{\frac{2}{3}} C \right)^2 \geq 0 \mbox{ and } \left( \frac{1}{\sqrt{6}} B + \sqrt{\frac{2}{3}} C \right)^2 \geq 0 \, \]
we also have
\begin{eqnarray*}
2AB + 2AC + 2BC &=& 2AB + \frac{4}{3} BC + 2 AC + \frac{2}{3} BC \\
&\geq & -\left( \frac{3}{2} A^2 + \frac{2}{3} B^2 + \frac{2}{3} C^2 \right) - \left( \frac{1}{6} B^2 + \frac{2}{3} C^2 \right) \\
&\geq & -\left( \frac{3}{2} A^2 + \frac{5}{6} B^2 + \frac{4}{3} C^2 \right) \, .
\end{eqnarray*}
Taking $A= r\tilde{\phi}_t$, $B= r\tilde{\phi}_r$ and $C= \tilde{\phi}$, this proves equivalence (\ref{ControlEMorawetz}). \qed


\begin{thebibliography}{100}

\bibitem{BaSeZo} J.C. Baez, I.E. Segal \& Z.F. Zhou, {\em The
global Goursat problem and scattering for nonlinear wave equations},
J. Funct. Anal. {\bf 93} (1990), p. 239-269.

\bibitem{ChriKla} D. Christodoulou \& S. Klainerman, {\em The
global nonlinear stability of the Minkowski space}, Princeton
Mathematical Series 41, Princeton University Press 1993.

\bibitem{ChruDe2002} P. Chrusciel \& E. Delay, {\em Existence of non trivial,
    asymptotically vacuum, asymptotically simple space-times},
    Class. Quantum Grav. {\bf 19} (2002), L71-L79, erratum
    Class. Quantum Grav. {\bf 19} (2002), 3389.

\bibitem{ChruDe2003} P. Chrusciel \& E. Delay, {\em On mapping properties of
    the general relativistic constraints operator in weighted function
    spaces, with applications}, preprint Tours Univervity, 2003.

\bibitem{Co2000} J. Corvino, {\em Scalar curvature deformation and a
    gluing construction for the Einstein constraint equations},
    Comm. Math. Phys. {\bf 214} (2000), 137--189.

\bibitem{CoScho2003} J. Corvino \& R.M. Schoen, {\em On the asymptotics
    for the vacuum Einstein constraint equations}, gr-qc 0301071,
  2003.

\bibitem{DaRo} M. Dafermos \& I. Rodnianski, {\em The redshift effect and radiation decay on black hole space-times}, arXiv gr-qc:/0512119, 2006.

\bibitem{Fri1980} F.G. Friedlander, {\em Radiation fields and
  hyperbolic scattering theory}, Math. Proc. Camb. Phil. Soc. {\bf 88}
  (1980), 483-515.

\bibitem{Fri2001} F.G. Friedlander, {\em Notes on the wave equation on
  asymptotically Euclidean manifolds}, J. Functional Anal. {\bf 184}
  (2001), 1-18.

\bibitem{HFri2004} H. Friedrich, {\em Smoothness at null infinity and the structure of initial data}, in The Einstein equations and the large scale behavior of gravitational fields, p. 121--203, Ed. P. Chrusciel and H. Friedrich, Birkhäuser, Basel, 2004.

\bibitem{KlaNi} S. Klainerman \& F. Nicol{\`o}, {\em On local and
global aspects of the Cauchy problem in general relativity},
Class. Quantum Grav. {\bf 16} (1999), p. R73-R157.

\bibitem{KlaNi2002} S. Klainerman \& F. Nicol{\`o}, The Evolution Problem
      in General Relativity, Progress in Mathematical Physics Vol. 25
      (2002), Birkha\"user.

\bibitem{KlaNi2003} S. Klainerman \& F. Nicol{\`o}, {\em Peeling properties of asymptotically flat solutions to the Einstein vacuum equations}, Class. Quantum Grav. {\bf 20} (2003), p. 3215-3257.

\bibitem{MaN04} L.J. Mason, \& J.-P. Nicolas, {\em Conformal scattering and the Goursat problem}, Journal of Hyperbolic Differential Equations, {\bf 1} (2) (2004), p. 197--233.

\bibitem{Mo1962} C.S. Morawetz, {\em The decay of solutions of the exterior initial-boundary value problem for the wave equation}, Comm. Pure Appl. Math. {\bf 14} (1961), p. 561--568.

\bibitem{Pe63} R. Penrose, {\em Null hypersurface initial data for classical fields of arbitrary spin and for general relativity}, in Aerospace Research Laboratories report 63-56 (P.G. Bergmann), 1963. Reprinted (1980)  in Gen. Rel. Grav. {\bf 12}, 225-64.

\bibitem{Pe65} R. Penrose, {\em Zero rest-mass fields including gravitation~: asymptotic behavior}, Proc. Roy. Soc. {\bf A284} (1965), p. 159--203.

\bibitem{PeRi84} R. Penrose and W. Rindler, Spinors and space-time, Vol. I (1984) and Vol. 2 (1986), Cambridge University Press.

\end{thebibliography}
\end{document}